\documentclass{aa}
\usepackage{graphics}
\usepackage{graphicx}
\usepackage{amssymb}

\begin{document}

   \thesaurus{05(08.08.1, 08.11.1, 10.05.1, 10.11.1, 10.15.1, 10.15.2 HR 1614) }

   \title{The reality of old moving groups - the case of HR 1614
   \thanks{Based on observations with the ESA Hipparcos satellite} }

   \subtitle{Age, metallicity, and a new extended sample}

   \author{Sofia Feltzing and Johan Holmberg}

   \offprints{Sofia Feltzing, email: sofia@astro.lu.se}
  \institute{Lund Observatory, Box 43, SE-221 00 Lund, Sweden
   }
 
%   \authorrunninghead{}
%   \titlerunninghead{}

   \date{Received 20 September 1999; accepted 7 March 2000}

   \maketitle

   \begin{abstract} We prove the existence of the old and metal-rich
moving group HR 1614. This is done using the new Hipparcos parallaxes
in combination with metallicities  derived from Str\"omgren
photometry, supported by  dynamical simulations of the evolution of old
moving groups in the galactic  potential.  A new selection criterion
for  this moving group is presented as well as a new, extended sample
of probable member stars.

 In particular we find that the HR 1614 moving group has an age of
about 2 Gyr (using Bertelli et al. 1994 isochrones) and a [Fe/H]
$\simeq 0.19 \pm 0.06$ dex.

We also revisit and apply our new selection criterion to  the samples
in Eggen (1992) and Eggen (1998b). It is found that, when
binaries and stars with too low metallicity have been removed, 15 of
his stars fulfill our criteria.

\keywords{Stars: Hertzprung-Russell and C-M diagrams, kinematics, 
Galaxy: evolution, kinematics and dynamics, open clusters and associations: 
general, open cluster and associations: individual: HR 1614 moving group}

\end{abstract}
%
%________________________________________________________________

\section{Introduction}

Globular and open clusters provide useful probes of the longterm
chemical and  dynamical evolution of the Milky Way. The globular
clusters probe the formation and early evolution of the spheroidal
components of the Milky Way while the open clusters provide a useful
tool to study the evolution of the galactic disk.  However, the
paucity of very old (old is here taken to mean $\gtrsim 10^8$ years,
the time scale on which an open cluster will be dissolved, Spitzer
1958) open clusters in the disk  forces us to consider the
considerably more loosely arranged moving groups to probe the earlier
evolution of the disk. It was Olin Eggen who first introduced the
concept of moving or stellar kinematic groups, of which
the Hyades is a well known example. The basic idea behind the moving groups 
is that  stars
form in clusters and thus with similar space motion, on top of which
the random motions of  single stars are added, resulting in a modest
velocity dispersion  within the group. Through the orbital motion
within  the galactic potential the group will be stretched out into a
tube-like structure and  finally, after several galactic orbits,
dissolve. The result of the stretching is that the stars will appear,
if the Sun happens to be inside the tube, all over the sky but may be
identified as a group through their common space velocity.  Thus the
moving groups may provide the essential, and so far largely
un-utilized, link between cluster and field stars.
These are the assumptions, but are moving groups observable 
realities? A large stumbling
block for assessing the reality of moving groups has been the lack of
large numbers of reliable parallaxes.  This has now been largely
overcome by the observations from the Hipparcos satellite (ESA
1997). This has, in fact, resulted in a small burst of recent papers
studying, mainly young, moving groups, e.g.  Asiain et al. (1999),
Barrado y Navascu\'es (1998), Odenkirchen et al. (1998),
Skuljan et al. (1997), and Dehnen (1998). 

Eggen defined moving groups as stars that all share the same velocity
in the direction of galactic rotation, i.e. V-velocity.
Specifically
the velocities required to be constant were corrected for the stars
differing radial distance from the Sun in order to make 
the circular orbits iso-periodic (Eggen 1998b).
 However,
using the Hipparcos parallaxes it is noted that firstly the groups get
more compact and secondly that stars identified as group members do
not form flat bars or ellipses with small $\sigma_V$ but are in fact
structures tilted in the $UV$-plane (Skuljan et al. 1997 Fig. 1). Part
of this shape can be attributed to the errors in the parallaxes
themselves and their transformation into errors in the
$UV$-plane. However, through dynamical simulations Skuljan et al. (1997)
show that all of the tilt cannot  be attributed to the errors in
parallaxes but also has a physical basis.

In view of these new possibilities  it is now appropriate to re-asses
the reality and membership criteria  for the HR 1614 moving group.
Eggen (1998b) has compared Hipparcos and cluster parallaxes for stars
in his sample (Eggen 1992) of HR 1614 moving group member
stars. However, he disregards the Hipparcos parallaxes in favour of
cluster parallaxes, also  when the discrepancies are large, without
further discussion. 

The article is organized as follows: Sects. 2, 3, and 4
describe the search for the HR 1614 moving group in the Hipparcos
catalogue, as well as dynamical simulations of old moving groups
and their characteristics today. 
Sect. 5 reviews previous work on the HR 1614 moving group.
A new selection criterion for HR 1614 moving group is developed in Sect.
6 and used to derive its age. In Sect. 7. we derive, from data in
the literature, a metallicity for the moving group. The Eggen (1992,
1998b) sample is revisited in Sect. 8 and discussed in detail. Sect. 9
contains a discussion primarily of possible sources of contamination
in our sample. Sect. 10 provides a brief summary of the main results of this 
paper.

\section{Finding the HR 1614 moving group -- 
Searching the Hipparcos catalogue}
\label{search.sect}

\begin{figure}
\resizebox{\hsize}{!}{\includegraphics{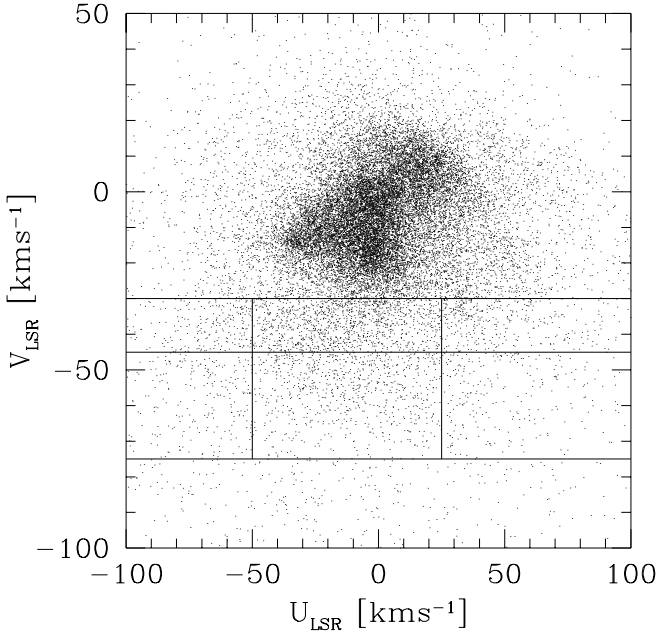}}
\caption[]{A relevant cut-out of the $UV$-plane for all stars 
in the Hipparcos
catalogue  with measured radial 
velocities (see the text) and $\sigma_{\pi} / \pi < 0.125$. 
The seven boxes investigated are outlined with solid lines.}
\label{uv.hip.fig}
\end{figure}

\begin{figure*}
\resizebox{\hsize}{!}{\includegraphics{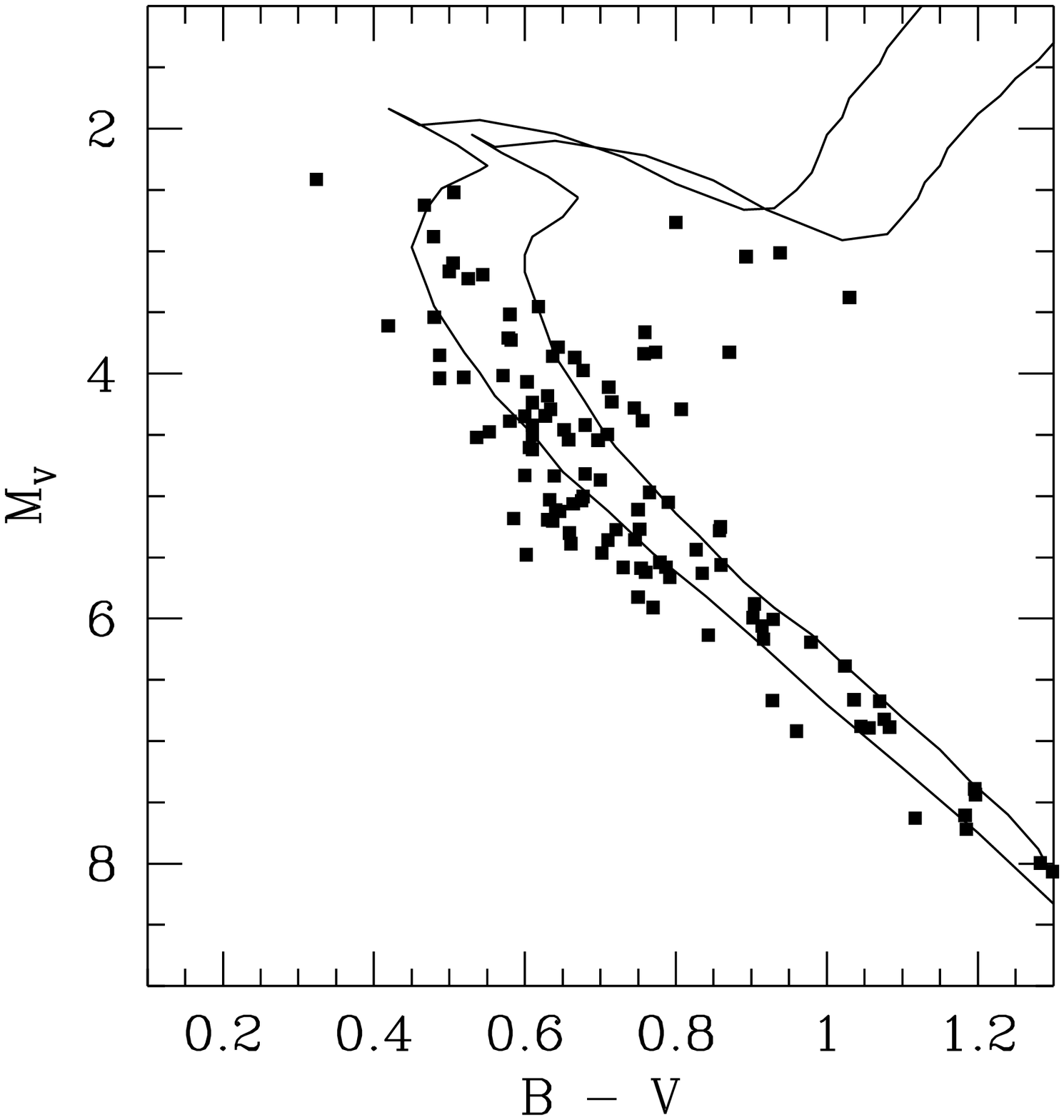}
\includegraphics{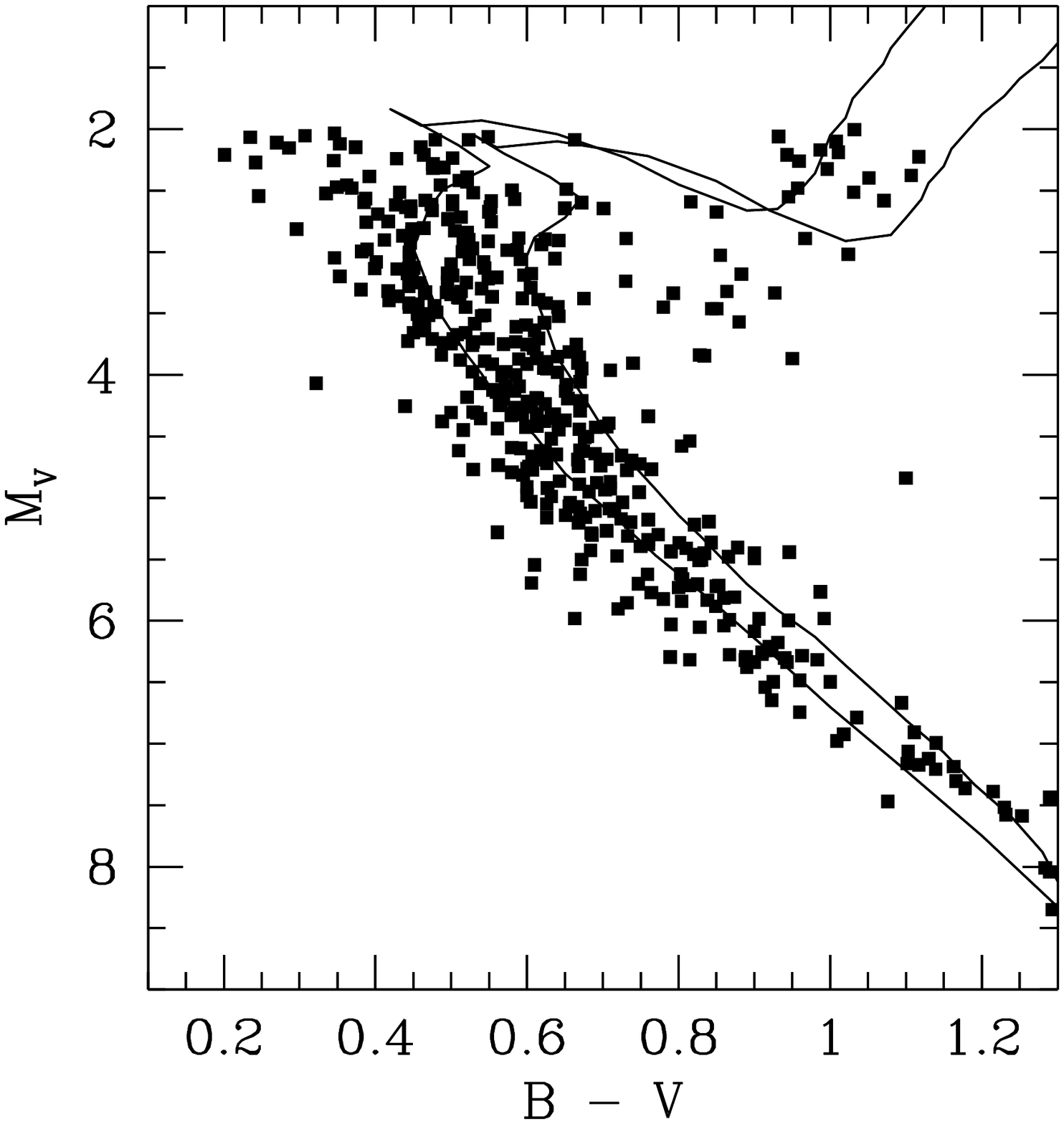}\includegraphics{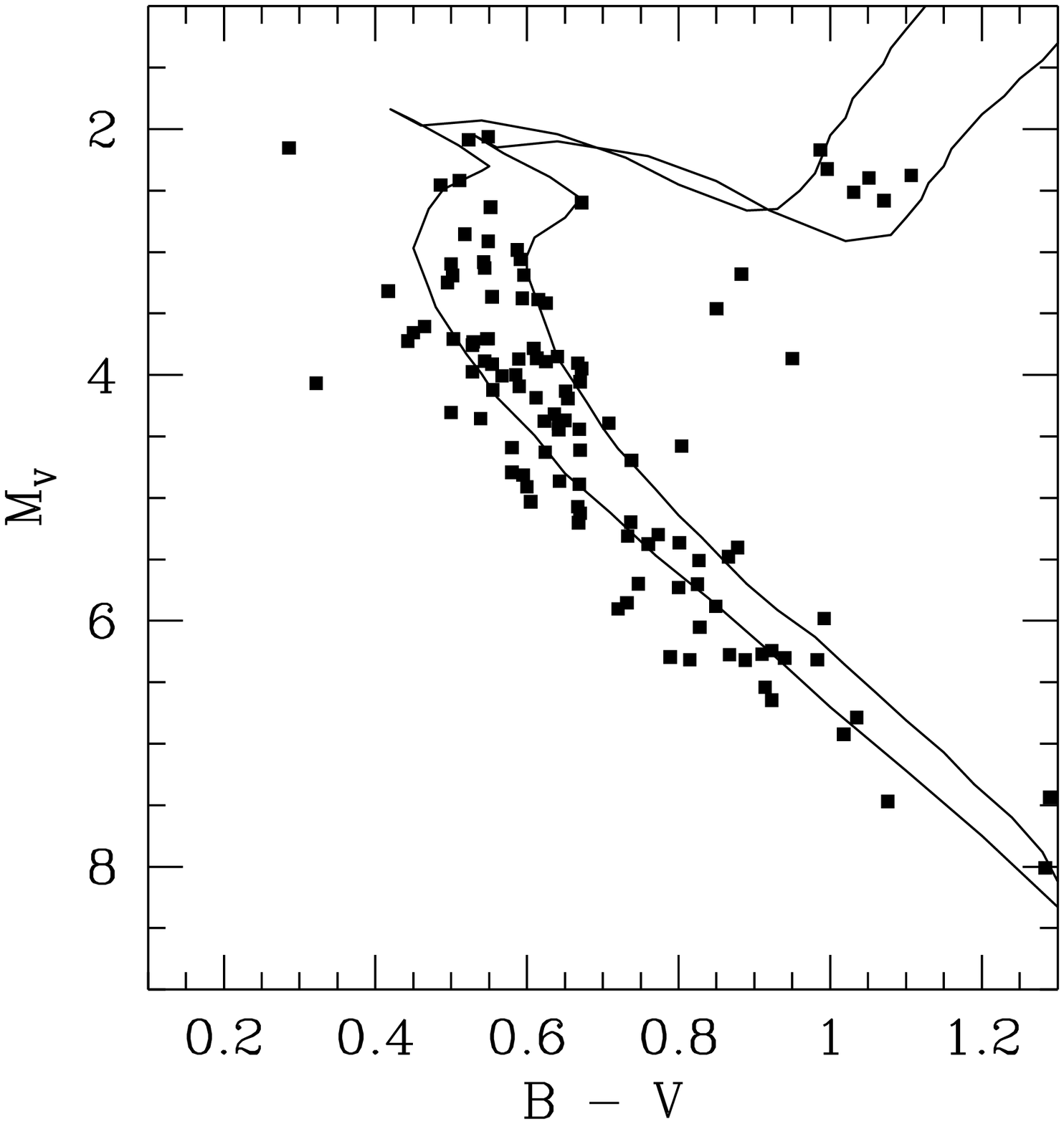}}
\resizebox{\hsize}{!}{\includegraphics{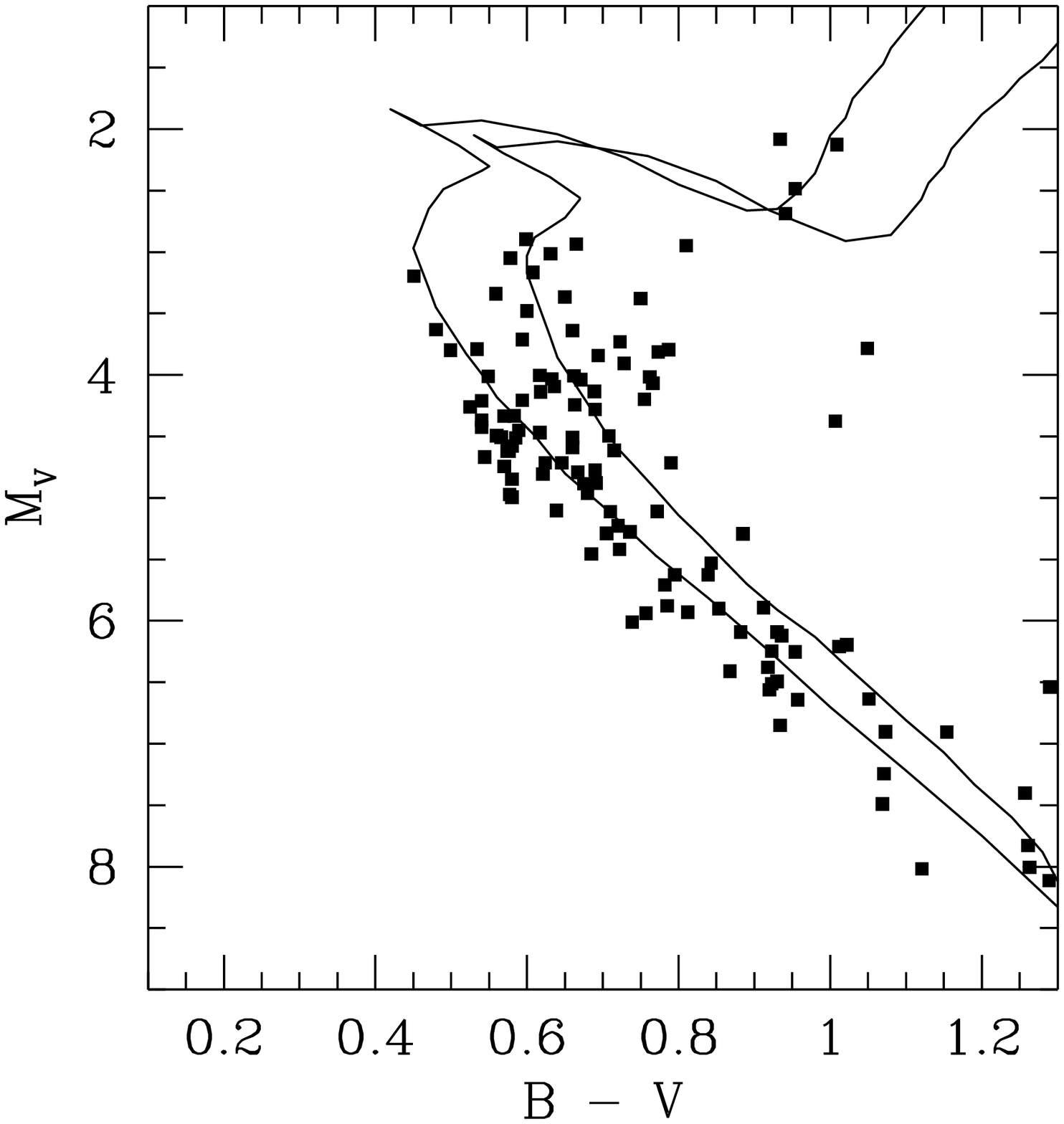}
\includegraphics{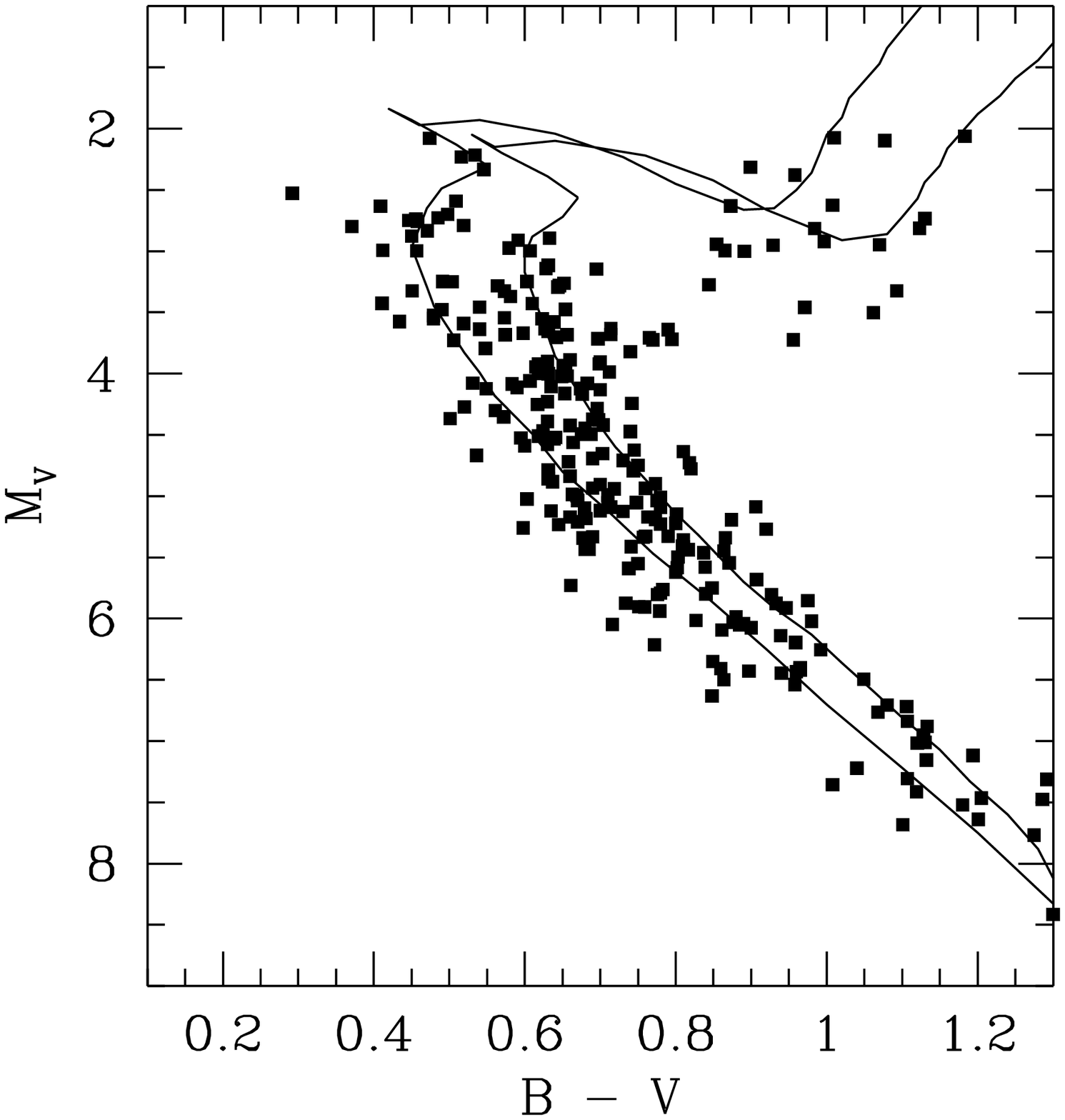}\includegraphics{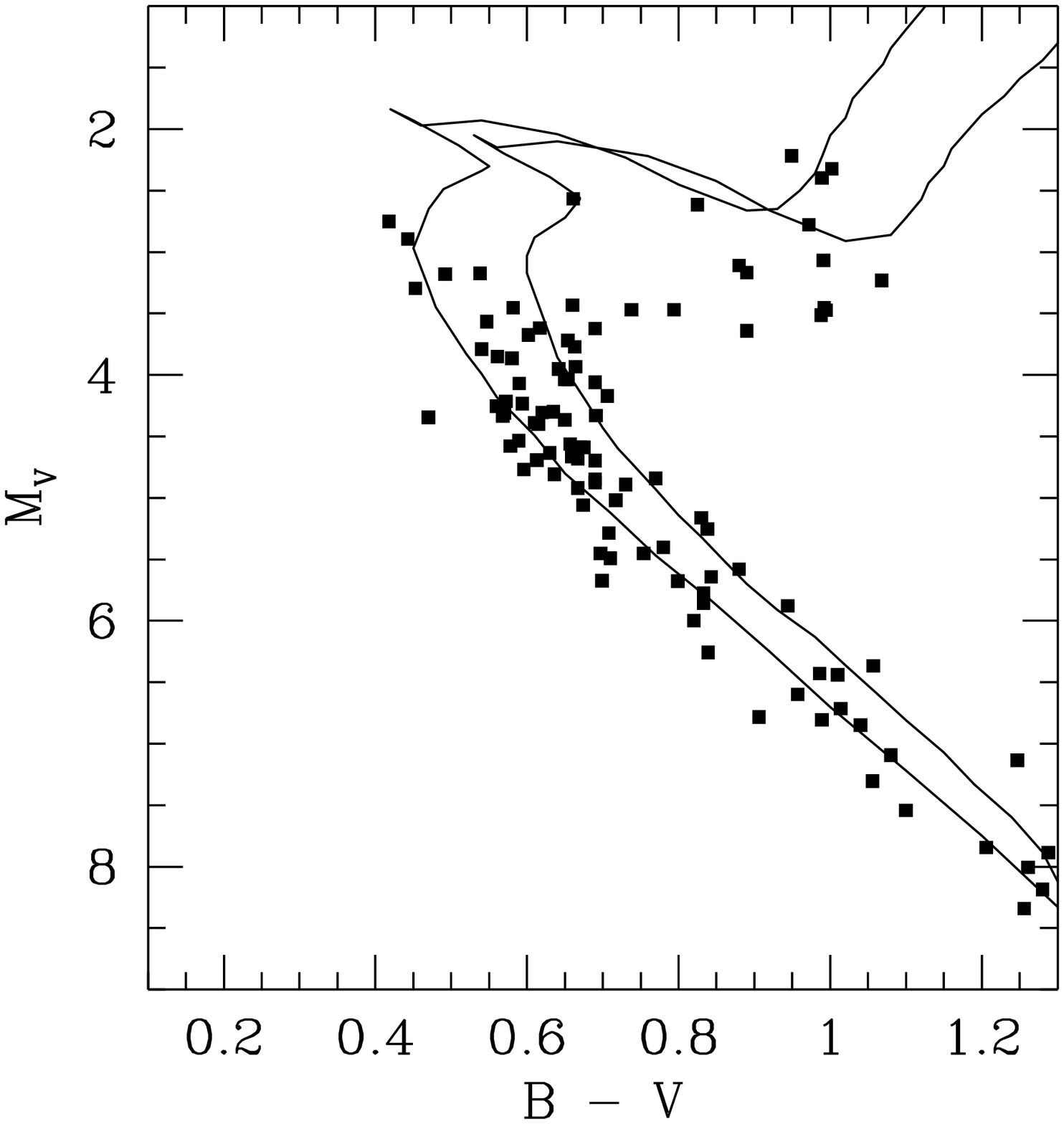}}
\resizebox{12cm}{!}{\includegraphics[width=3cm]{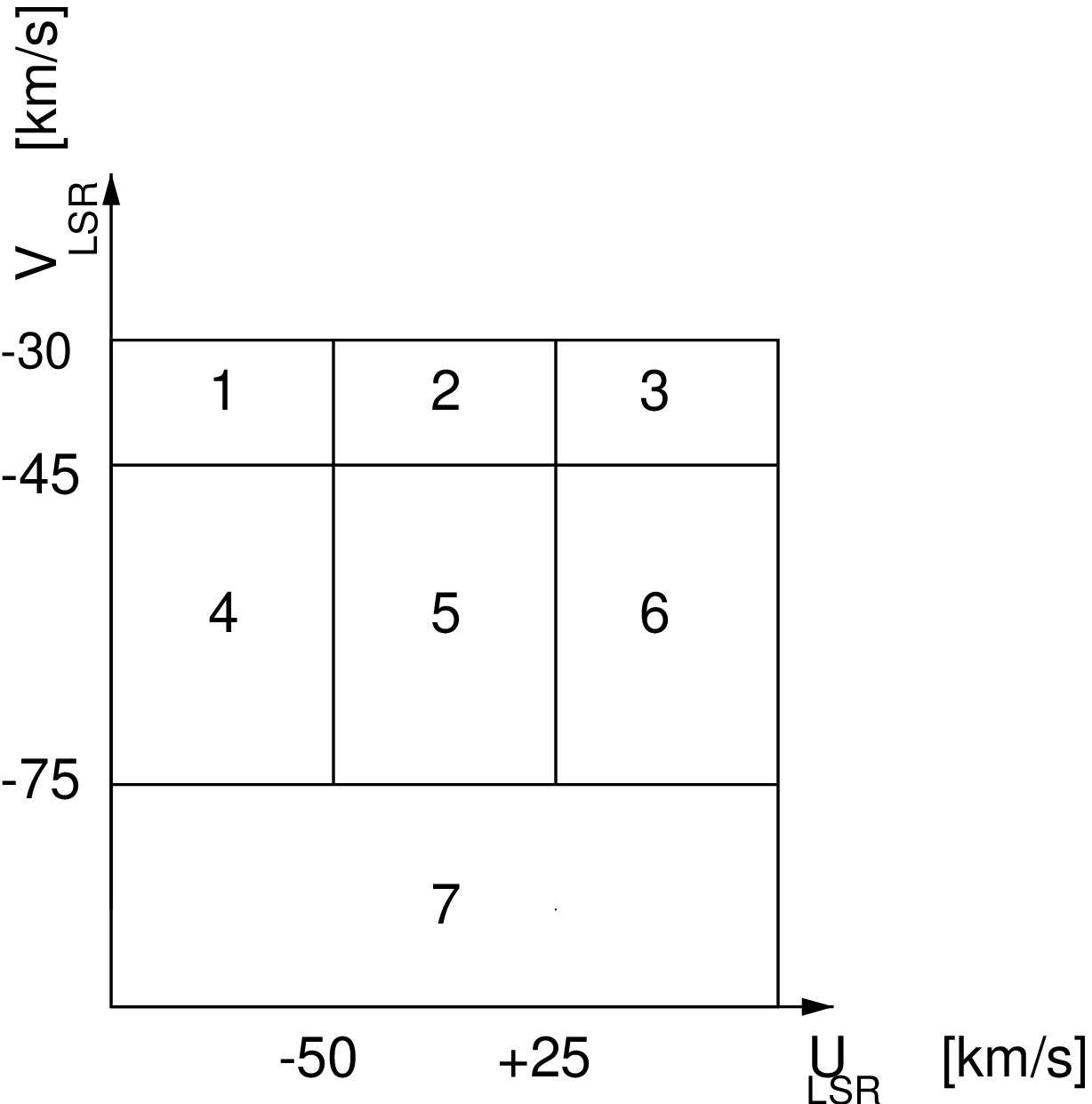}
\includegraphics[width=3cm]{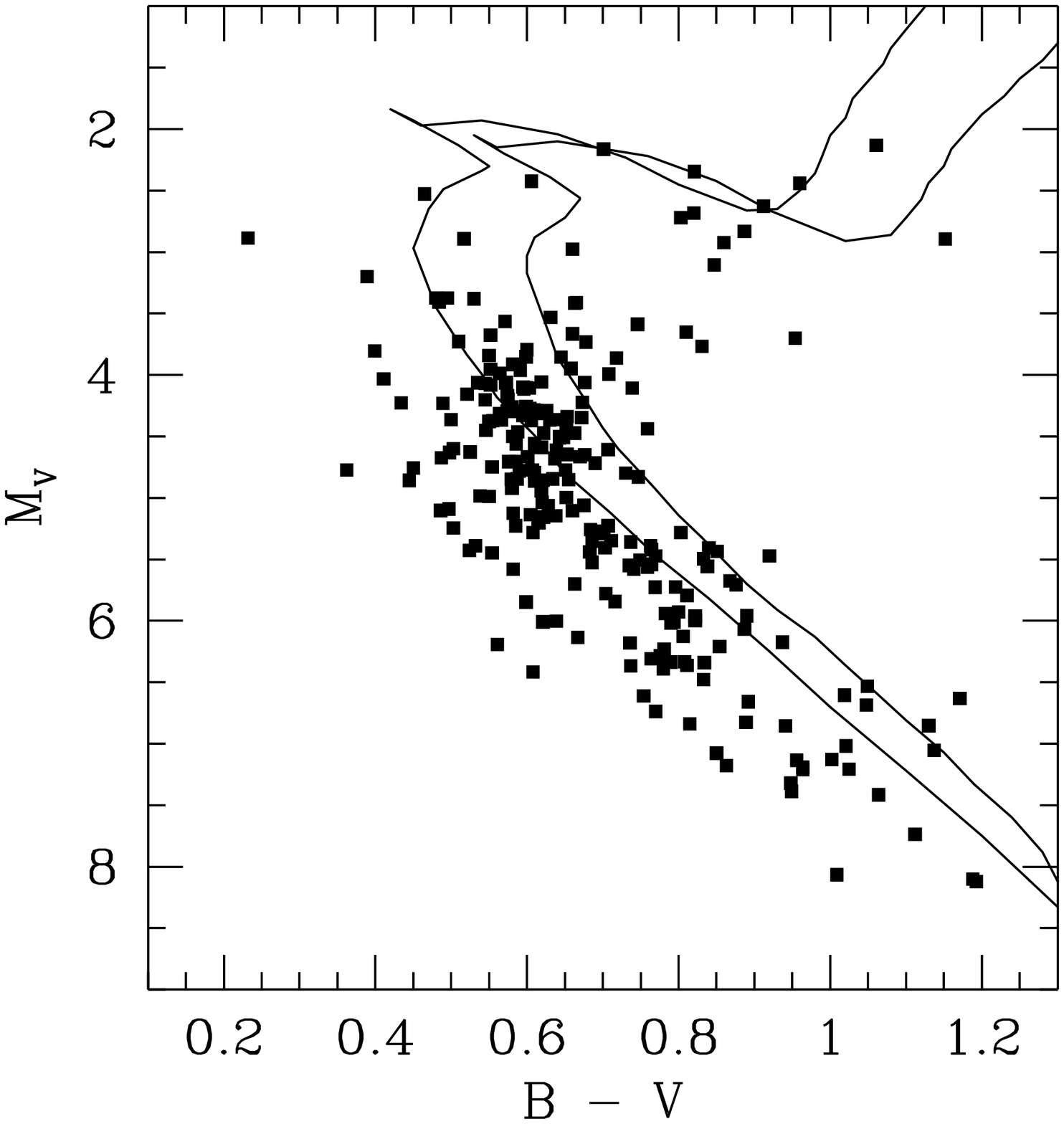}}
\caption[]{HR-diagrams for the seven boxes defined in Fig. \ref{uv.hip.fig}.
To help guide the eye two isochrones of 2 Gyr and Z=0.02 and 0.05
 respectively (corresponding to  [Me/H] = 0.0 and 0.4 dex)  from
Bertelli et al. (1994) are also plotted. No stars brighter than  $M_V
=2 $ have been plotted and  $\sigma_{\pi} / \pi < 0.125$ for all stars. }
\label{hrboxes.fig}
\end{figure*}

\begin{figure*}
\resizebox{12cm}{!}{\includegraphics{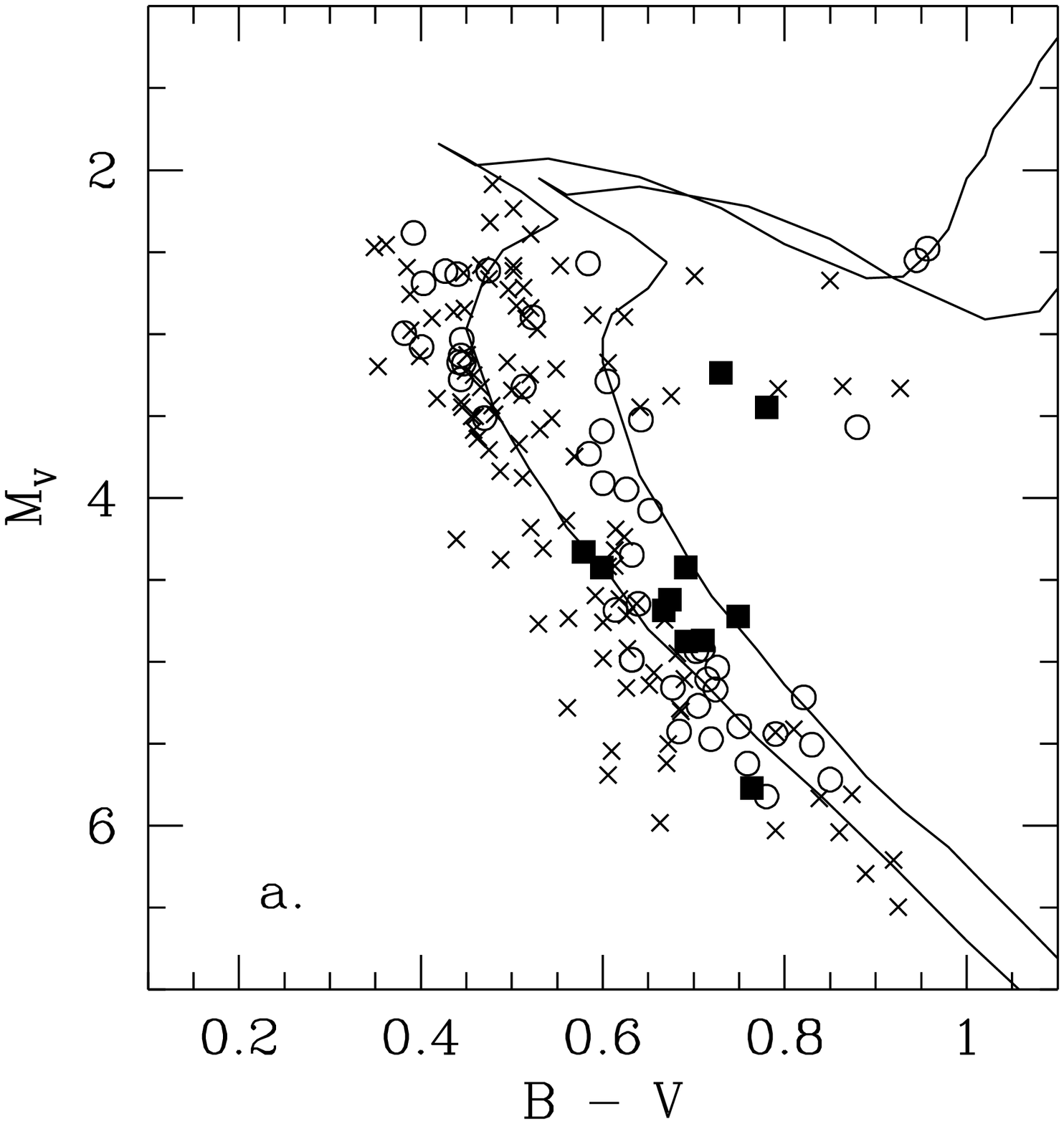}
\includegraphics{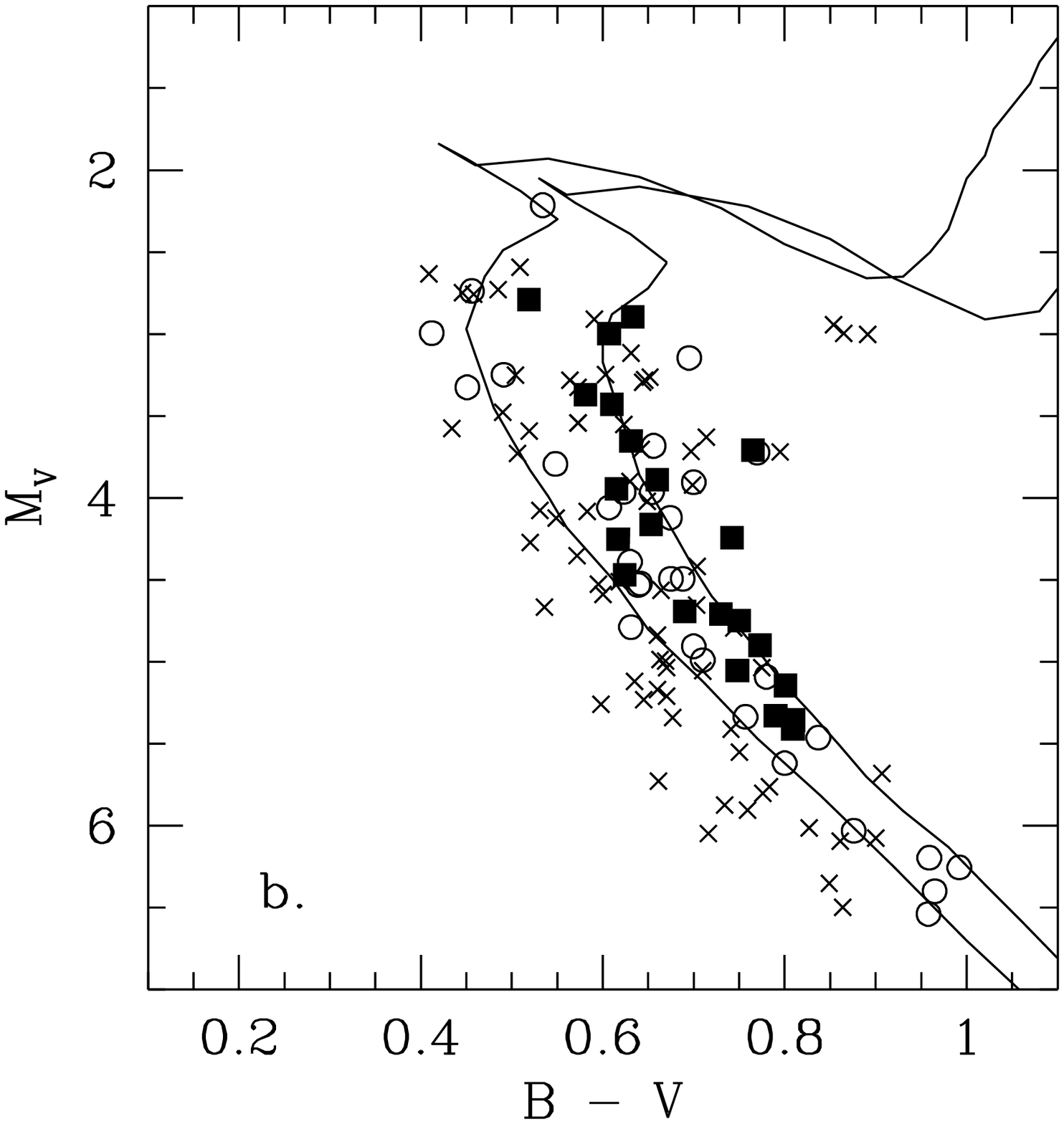}}
\hfill
\parbox[b]{55mm}{
\caption[]{ HR-diagrams for box 2 (a.) and  5 (b.)
 now only with single stars with available Str\"omgren
photometry and [Me/H] calculated from the calibration in  Schuster \&
Nissen (1989). $\times$ denote stars with [Me/H] $\leq -0.1$ dex, $\circ$
stars with $-0.1<$ [Me/H] $\leq 0.1$ dex, and  $ \blacksquare $ stars with
[Me/H] $>0.1$ dex. To help guide the eye two isochrones of 2 Gyr and Z=0.02 and 0.05
 respectively (corresponding to  [Me/H] = 0.0 and 0.4 dex)  from
Bertelli et al. (1994) are also plotted. 
No stars with absolute magnitude brighter than 2 are
plotted and  $\sigma_{\pi} / \pi  < 0.125$ for all stars.}
\label{hrboxesstr.fig}}
\end{figure*}

We perform an unbiased search in the Hipparcos catalogue complemented
with radial velocities from the Hipparcos Input 
Catalogue (ESA 1992), Grenier et al. (1999), and 
Barbier-Brossat et al. (1994) over a wide
area in $UV$-space (Fig. \ref{uv.hip.fig} shows the distribution  of
stars) around the probable values for the HR 1614 moving group to see
if we can find any signature of what appears to be stars with similar,
high, metallicities and with correlation in the $UV$-plane.   First we
divide the $UV$-plane into seven boxes and construct the corresponding
HR-diagrams, Figs. \ref{uv.hip.fig} and \ref{hrboxes.fig}.

The four  boxes 1, 3, 4, and 6, all show 
stellar population with metallicities around solar and below. Box 3
has  what appears to be a younger population as well, while the other
boxes show exclusively old populations. None of these boxes show any
trace of a large  population significantly more metal-rich than the
Sun and we will disregard them from further discussions and
concentrate on the remaining three boxes.

The three boxes 2, 5, and 7 all show well populated HR-diagrams. As
we move  from box 2 to 7 over box 5 the turn-off age of stars with
solar-like metallicities increases. 
The canonical view of the stellar populations in the solar neighbourhood 
implies that we would expect subsequently more and more metal-poor
stars as we move further  away from the Sun in velocity
space. Especially we would expect box 5 to be older and more
metal-poor than box 2 and box 7 even older and more metal-poor as
we sample more and more halo and thick disk stars and less of the disk stars. 

\begin{figure}
\resizebox{\hsize}{!}{\includegraphics[angle=-90]{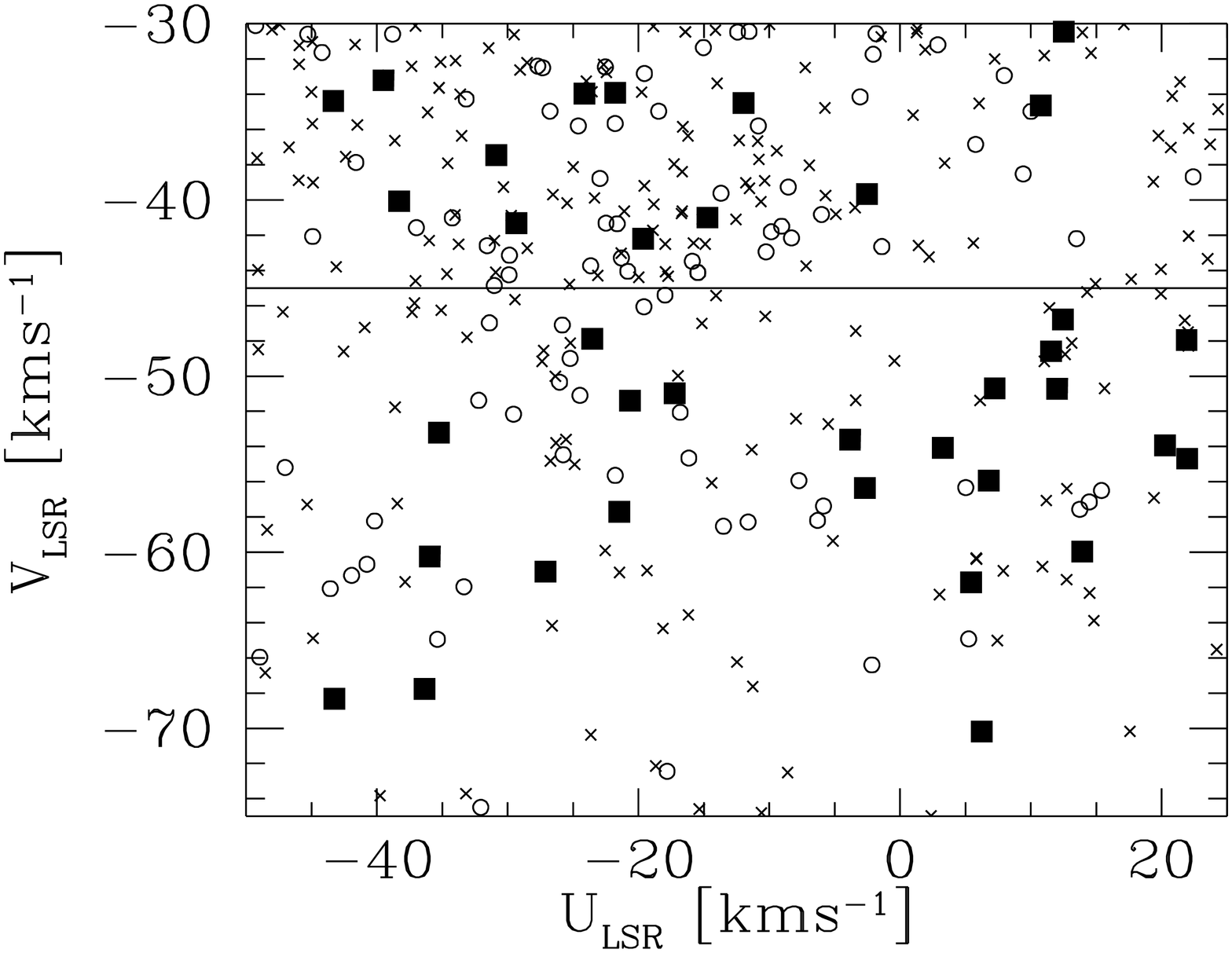}}
\caption[]{Distribution of the stars in the Str\"omgren catalogue
in the $UV$-plane with the same symbols as in Fig. \ref{hrboxesstr.fig}.
The dashed line represents the boundary between box 2 and 5, see Fig.
\ref{uv.hip.fig}. }
\label{uvbox25.fig}
\end{figure}

\begin{figure}
\resizebox{\hsize}{!}{\includegraphics[angle=-90]{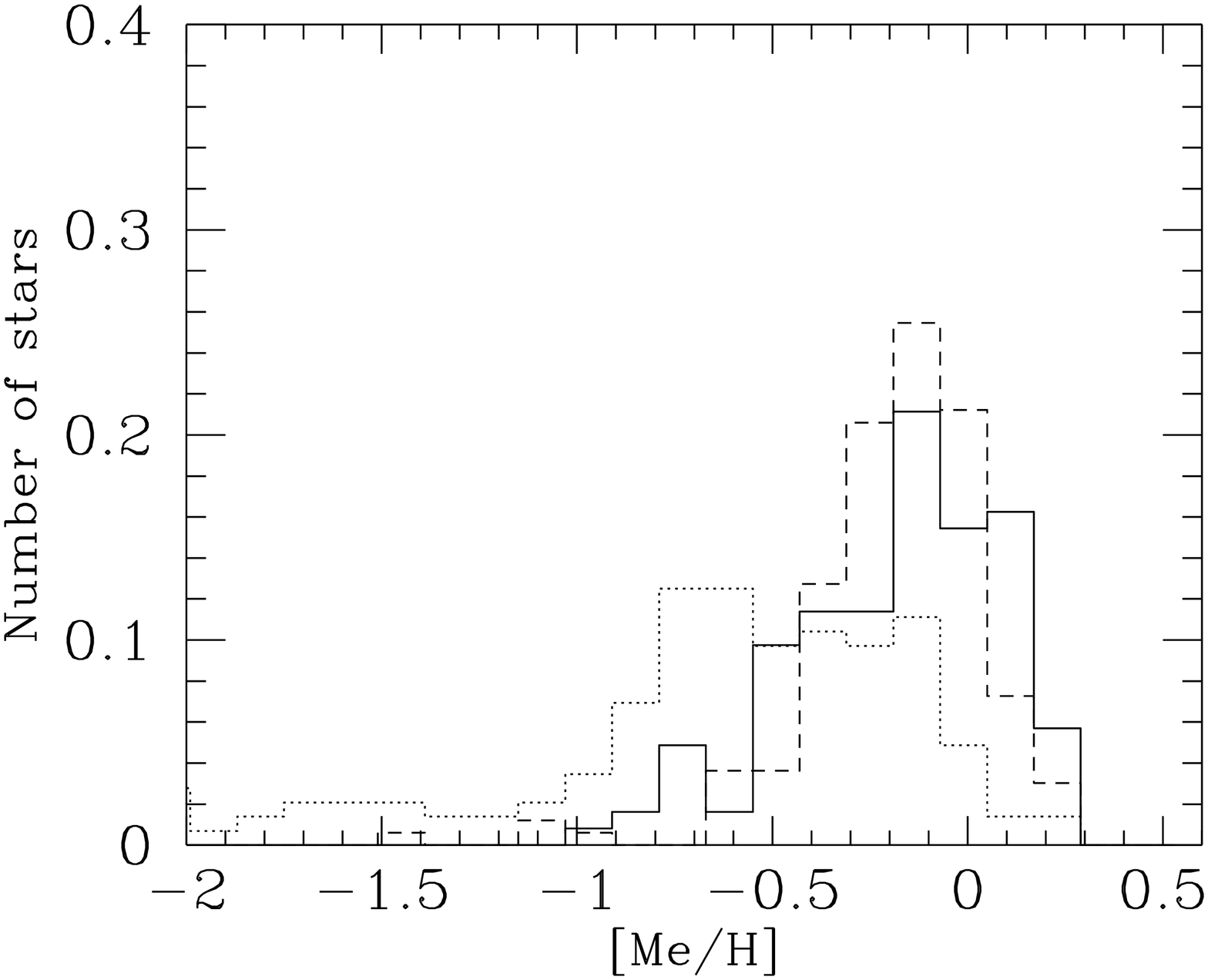}}
\caption[]{The metallicity distributions in box 2 (dashed line),  box
5 (full line), and box 7 (dotted line). No stars brighter than
$M_V=2$  are included and all stars have  $\sigma_{\pi} / \pi < 0.125$.}
\label{methist.fig}
\end{figure}

The visual impression from Fig. \ref{hrboxes.fig} is the presence of a 
a richly populated metal-rich isochrone present in box 2 and 5
but not in any other box. However the relative number of 
metal-rich stars appears much larger in box 5 than in box 2. 
Is this significant? To find out
we combine our
Hipparcos catalogue with  Str\"omgren photometry (Hauk \& Mermilliod
1998) and calculate [Me/H] using the calibration in Schuster \& Nissen
(1989).  
All stars flagged as possible binaries in the Hipparcos catalogue
were excluded. This includes both systems detected by Hipparcos itself
and system previously known (e.g. from radial velocity variation) and 
included in the CCDM catalogue of multiple stars (ESA 1997).
 These stars have to be excluded 
 since it is not possible to derive metallicities from their 
Str\"omgren photometry. The results are shown in Figs. \ref{hrboxesstr.fig} 
and \ref{uvbox25.fig}
where we also distinguish between three
major metallicity ranges, [Me/H] $< -0.1$, $-0.1 \leq $[Me/H]$ <
+0.1$, and $+0.1 \leq$ [Me/H].  Again box 5 shows a clear high-metallicity
population not at all present in the other field.  The resulting
normalized metallicity-distributions for box 2, 5, and 7 are shown in
Fig. \ref{methist.fig}. Clearly box 7 has a large,
metal-poor tail, as expected since this box should contain many halo
stars as well as thick disk stars.  Box 2 on the other hand, which
should be the most  solar-like box, has almost no metal-poor tail and
a distinct peak at $\sim -0.2$ dex, just what is expected  for the
solar neighbourhood, Wyse \& Gilmore (1995).  Box 5 on the other hand
is more metal-rich than the solar neighbourhood. This does not fit
into the canonical picture of the general stellar populations in 
the Galaxy. It should be more metal-poor (on average)  than box 2 and
more metal-rich than box 7.

The background distribution of stars in Fig. \ref{uv.hip.fig} shows
that the maximum density occurs in several clumps rather close to the
local standard of rest (LSR)  but also that a secondary maximum exists
centered on  $U\approx -20$ km~s$^{-1}$, $V\approx -40$ km~s$^{-1}$
(see also Dehnen 1999, Fig. 1 for a similar smoothed density
distribution plot). This stellar population with a mostly negative
U-velocity, was already tentatively identified in pre-Hipparcos data
and  named the U-anomaly. In the Hipparcos data it is much  more well
defined. Raboud et al. (1998) who obtained Geneva photometry for stars
in this area of the $UV$-space, found a metallicity distribution
similar to that of the bulge. Supported by dynamical simulations they
concluded that the stars formed in the inner disk were scattered by
the bar into the solar neighborhood. These stars make up a large part
of the old stars found in box 2. Figure  \ref{uvbox25.fig} shows the
velocity distribution of stars with Str\"omgren metallicities in box 2
and 5. The U-anomaly is clearly seen in box 2 and found to be
dominated by stars of solar metallicity and below.

From this and our previous considerations it is clear that the
super-solar metallicity stars are predominantly found in box 5, the
area of the $UV$-plane in which the HR 1614 moving group is supposed to be
found. We thus conclude  that there exists a distinct
stellar population more
metal-rich than the average background centered  at $U \sim -10$ and
$V \sim - 60$ km~s$^{-1}$ and tilted in  the $UV$-plane. This population
is not found in the adjacent areas in the $UV$-plane. Can this stellar
population be identified with the moving group HR 1614 found by Eggen?

\section{Finding the HR 1614 moving group -- Dynamic simulations}
\label{dynsim.sect}

To answer this question we now turn to a  dynamical simulation
of the evolution of old moving groups to find out whether or not 
the structure we observe in box 5 can indeed be identified with an old 
moving group.

\begin{figure*}
\resizebox{12cm}{!}{\includegraphics{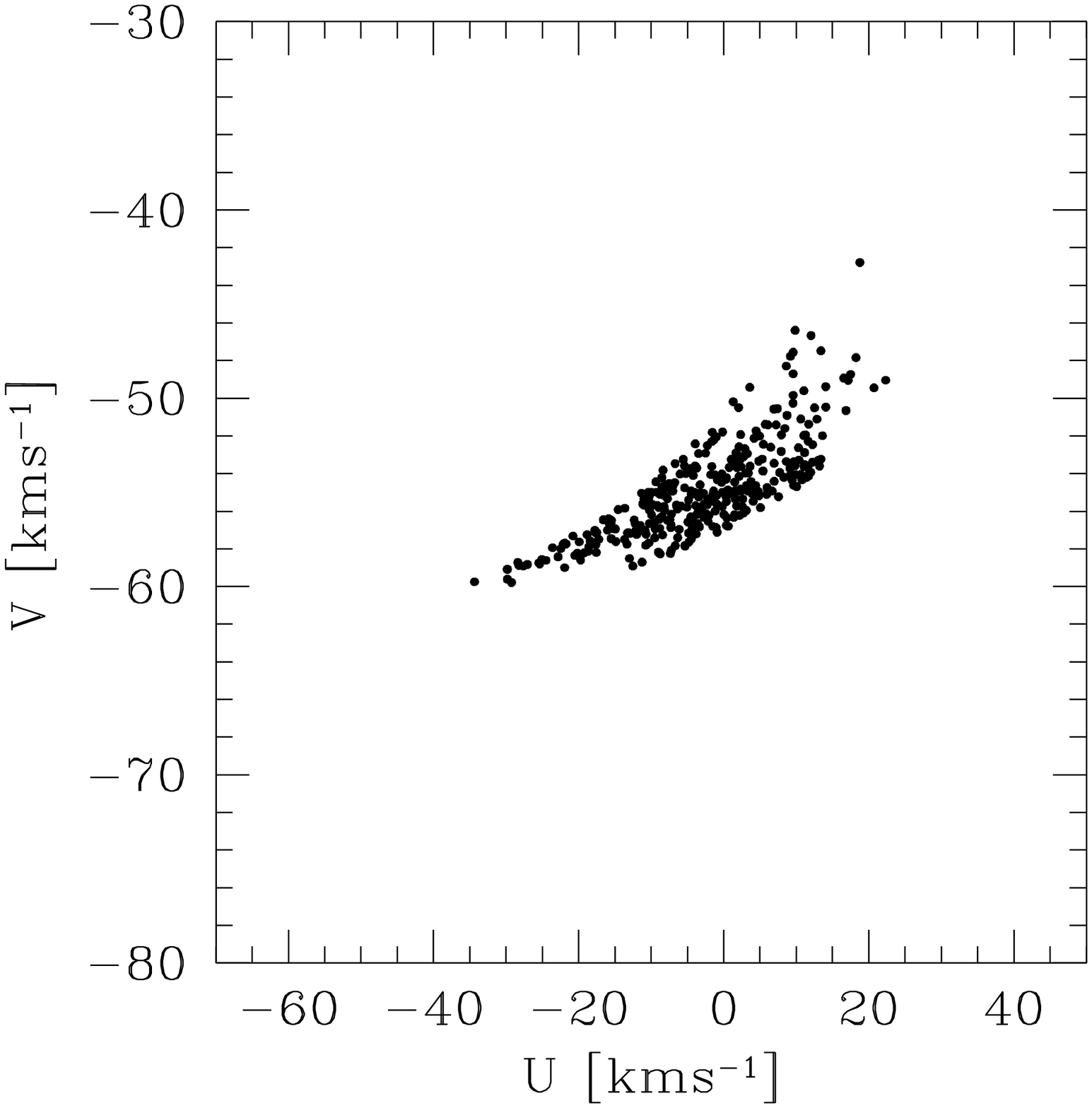}
\includegraphics{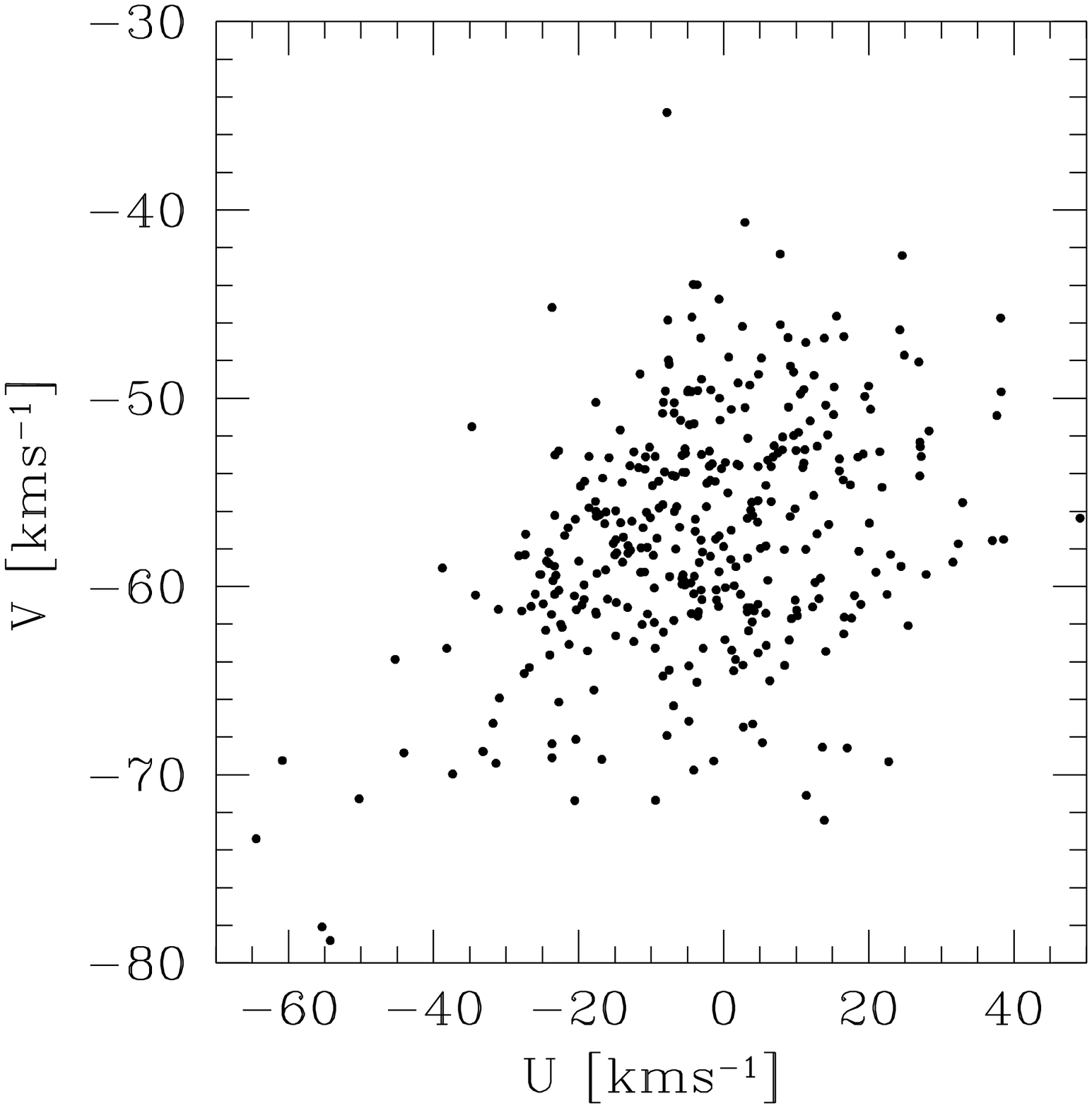}}
\hfill
\parbox[b]{55mm}{
\caption[]{Two typical outcomes from the dynamical 
simulations described in Sect. \ref{dynsim.sect}. {\bf a.} when 
the cluster is born with an internal velocity dispersion and then
no further processes increase the dispersion. {\bf b.} when Wielen
orbital diffusion is used to mimic the physical processes that might 
increase the scatter.}
\label{sim.fig}}
\end{figure*}

A galactic potential model consisting of a stellar and gaseous disk,
thick  disk, bulge and dark halo (Dehnen \& Binney 1998) is used to
integrate the orbits of the individual stars in a dissolving moving
group. In the model the solar distance from the galactic centre is
R$_{\odot}=8.0$ kpc, the height above the plane is z$_{\odot}=8$ pc,
the solar peculiar velocity components are $U=10$ km~s$^{-1}$, $ V=5$
km~s$^{-1}$, $W=7$ km~s$^{-1}$ and the circular  velocity
$v_{\rm c}(R_{\odot})=219$ km~s$^{-1}$. The starting point for the
simulation is attained by taking a velocity vector and position  from
a star  assumed to belong to the moving group e.g. HR 1614 itself and
integrate the  orbit backwards for the assumed age of the moving
group. At this position an  ensemble of stars is placed and followed
forward into the present time. The  stars are treated as test
particles in the static potential and the dispersion processes is
modeled in two different ways. Either the moving group is born
unbound like an OB association with a certain velocity dispersion and
no further dispersion processes are active or the group is born as a
bound system like an open cluster with identical velocities and the
dispersion is gradually built up by stars becoming unbound to the
cluster and starting to experience  orbital diffusion (Wielen 1977).

 Both methods are rather crude approximations to the real processes
that affect the evolution of a dissolving moving group, but they can
be seen as limiting cases to the real processes at work. Our main use
of the simulations is also to test the assumption that the sample of
stars identified as probable members due to metallicity and kinematics
have a common origin in time and space. The two examples in
Fig. \ref{sim.fig} shows that the structure in the  $UV$-plane outlined
by the metal-rich stars in Fig. \ref{uvbox25.fig}
with a width in U-velocity of 60 km~s$^{-1}$ and in $V$ of 20 km~s$^{-1}$
and not confined to a single V-velocity is a natural result of the
dynamically simulated dispersion processes. The classical
configuration, used by Eggen, with all stars belonging to the group lined 
up along a
single V-velocity only appears when the Sun is located very close to the
centre of the  group, which is not likely.
 Otherwise, either if the Sun is located to the
outside or the inside of the tube-orbit defined by the moving group,
we get the tilted structure in the $UV$-plane shown in Fig. \ref{sim.fig}. In
the simulation in Fig. \ref{sim.fig}a with an  original velocity dispersion
of 6 km~s$^{-1}$, the center of the group had a velocity of $U=-15$ 
km~s$^{-1}$, $V=-58$ km~s$^{-1}$ and $W=-7$ km~s$^{-1}$ when it passed the
present Solar position 40 Myrs ago. In the simulation with Wielen
diffusion, we used a time and velocity independent diffusion
coefficient $D=2.0\cdot 10^{-7}$ km~s$^{-1}$ yr$^{-1}$ and the center of the
group had a velocity of $U=0$ km~s$^{-1}$, $V=-58$  km~s$^{-1}$ and
$W=-7$ km~s$^{-1}$ when it passed the present Solar position only 5 Myrs
ago. These two examples have their peak density in the $UV$-plane at
approximately the same position as the observed distribution found in
Sect. \ref{search.sect}.

\section{Finding the HR 1614 moving group -- Conclusions}
\label{findconcl.sect}

We close the first part of our paper with a few remarks on the
reality of old moving groups.  Before the Hipparcos mission the
numbers of reliable parallaxes were too small to address the reality
of most proposed old moving groups successfully.  Recent studies
(e.g. Barrado y Navascu\'es 1998, and Skuljan et al. 1997) have shown
that  well known moving groups such as the Pleiades and the  Hyades
but  also several other  young moving groups, e.g. Castor ($200\pm100$
Myr, Barrado y Navascu\'es 1998), are well identified as physical
entities using the new data from the Hipparcos mission.

Are the proposed old moving groups a reality? We conclude
that at least one old moving group exists and that it's possible to
find other ones using  velocity information in combination with
metallicities  based on Str\"omgren photometry.  The study by Dehnen
(1998) further supports our findings. He recovered many maxima, using
a maximum likelihood solution,  in the velocity distribution of nearby
stars using the Hipparcos catalogue. Several of these are identifiable
with known moving groups. In particular he found several maxima that
exclusively contained red stars, indicating an old age. He identifies
one of these maxima  with the HR 1614 moving group.

However, HR 1614 might be a rather special case. It is particularly 
metal-rich compared to the majority of stars in the part of the $UV$-plane
it resides in. In other parts of the $UV$-plane (e.g. close to the local standard
of rest) the group would have been completely obscured by other metal-rich
stars. 

In summary, we conclude that at least one old moving group exists and
 it's possible  to find others using our simple method if they stand out in
 terms of metallicity and/or age from the ambient background of stars
 in their space of the $UV$-plane.

\section{Review of previous work on the HR 1614 moving group}

The HR 1614 moving group stands out among stellar moving groups in
terms of age and metallicity. The age and metallicity  have been
estimated to be 
roughly similar to that of the old open cluster NGC 6791, Hufnagel
\& Smith (1994) and Eggen (1998a). However, this is challenged 
by the recent determination 
by Chaboyer at al. (1999) who found NGC 6791 to have a metallicity of $+0.4$
dex and an age of 8 Gyr.

It was Eggen (1978) who, following leads in earlier studies by Eggen
(1971) and Hearnshaw (1974), first identified the presence of a moving
group associated with the K dwarf star HR 1614 (HD 32147) by studying a
sample of  stars which were selected as being within $\pm 10$ km~s$^{-1}$ of
the V-velocity of the star
HR 1614 (then estimated to $-58$ km~s$^{-1}$). He found that
in a sample of 44 stars with $(U,V) $\footnote{These
are the  $U$ and $V$ velocities in a right handed system. 
Eggen use a left handed system, however, we conform with
the now common practice and use a right handed system where
U points in the direction towards the galactic centre and $V$ in the
direction of galactic rotation.} $\approx (0,-60)$
km~s$^{-1}$ 60 \%  were over abundant with
respect to the Sun (based on a few spectroscopic studies, mainly Oinas
(1974) and the $m_1$ index). 
The stars appear as metal-rich as the Hyades in the
$M_V,R-I$ diagram. They 
also showed strong blanketing effects in $b-y$ (an excess of
$\sim 0.03$ compared to the Hyades group, which was utilized as a
selection criterion) and made up a colour magnitude diagram resembling
that of an old stellar cluster (Eggen 1978 Fig. 1a).

Smith (1983) subsequently obtained DDO photometry of 19 suggested member 
stars  and found  many of them to have enhanced
cyanogen bands similar to those  found in so called super-metal-rich
stars (SMR, see Taylor 1996 for a discussion of SMR stars). The
group also showed anomalously many CN-rich stars when compared 
with a random sample of stars.  Derived [Fe/H]
generally confirmed the high metallicity of the member stars,
however, two giant stars were found to be significantly lower in
metallicity casting some doubt on Eggen's (1978) $b-y$ criterion for
membership. A stricter application of membership criteria,
i.e. that the stars should show the same behaviour in both
UBV, DDO and $b-y$, showed that many of  Eggen's original candidates did
not belong to the group. Smith (1983) further  suggests that the
abundances and  kinematics of the HR 1614 moving group stars are consistent
with the abundance gradient observed in old open clusters (if it
formed at perigalacticon) and thus the lack of observations of similar
clusters would be due to that either they no longer exist or that
observations up to 1983 had not sampled the inner regions of the
Galaxy well enough. Utilizing the CN-enhancement Eggen (1992) found
a total of 39 main-sequence members of the HR 1614 moving group by
isolating them through their V-velocity.  All these dwarfs are
within 40 pc of the Sun. A further 19 red giants within 200 pc of
the Sun were also identified.

No high resolution spectroscopic study has targeted the probable
member stars of the HR 1614 moving group.

\section{A new selection of probable HR 1614 members and its age}
\label{specab.sect}

We now proceed
with the second part of this paper and provide a new sample of stars
with high probability of being member stars. From this new sample  we
are able to derive an age as well as a metallicity for the moving
group. 

\begin{figure*}
\resizebox{12cm}{!}{\includegraphics{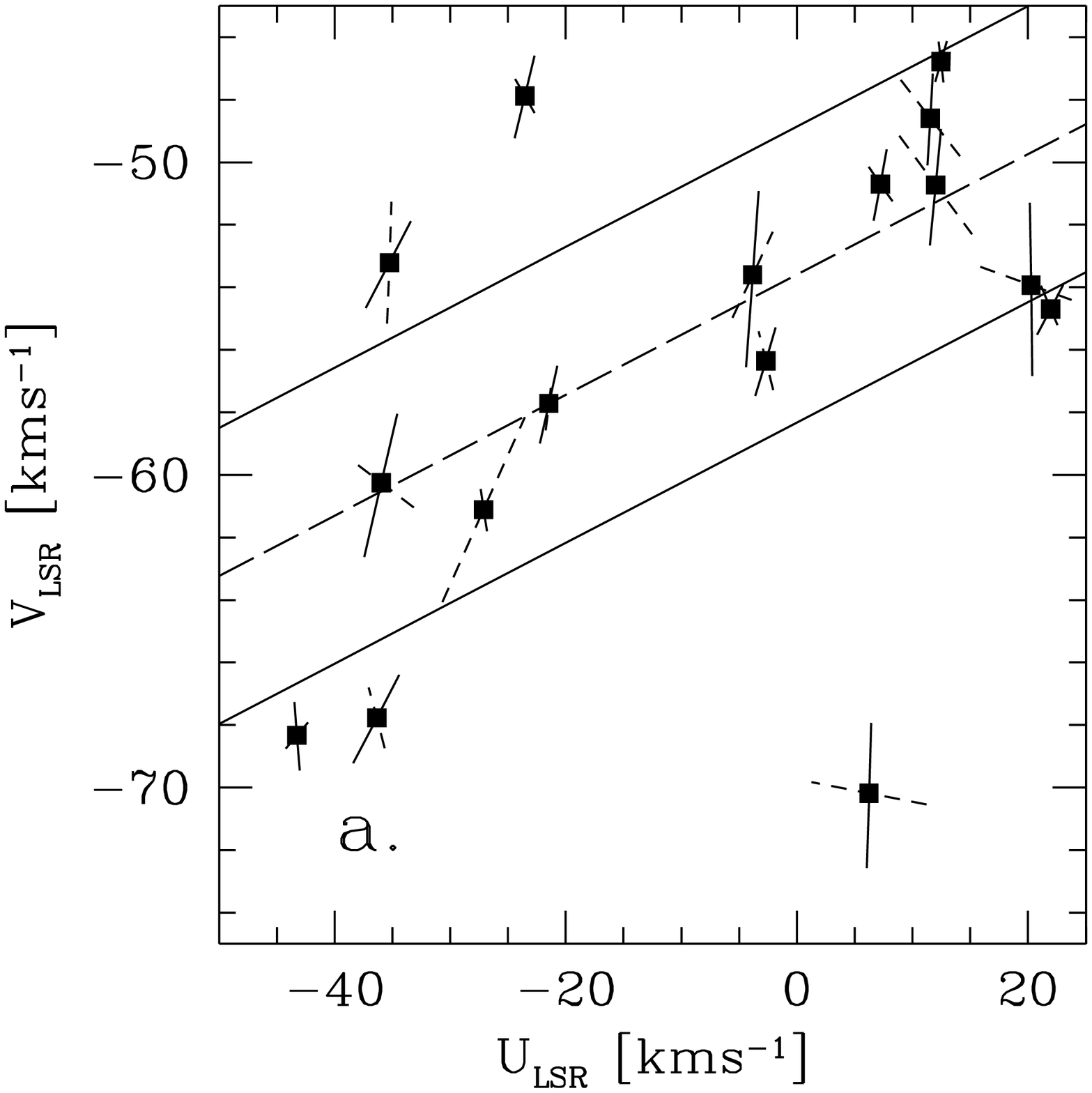}
\includegraphics{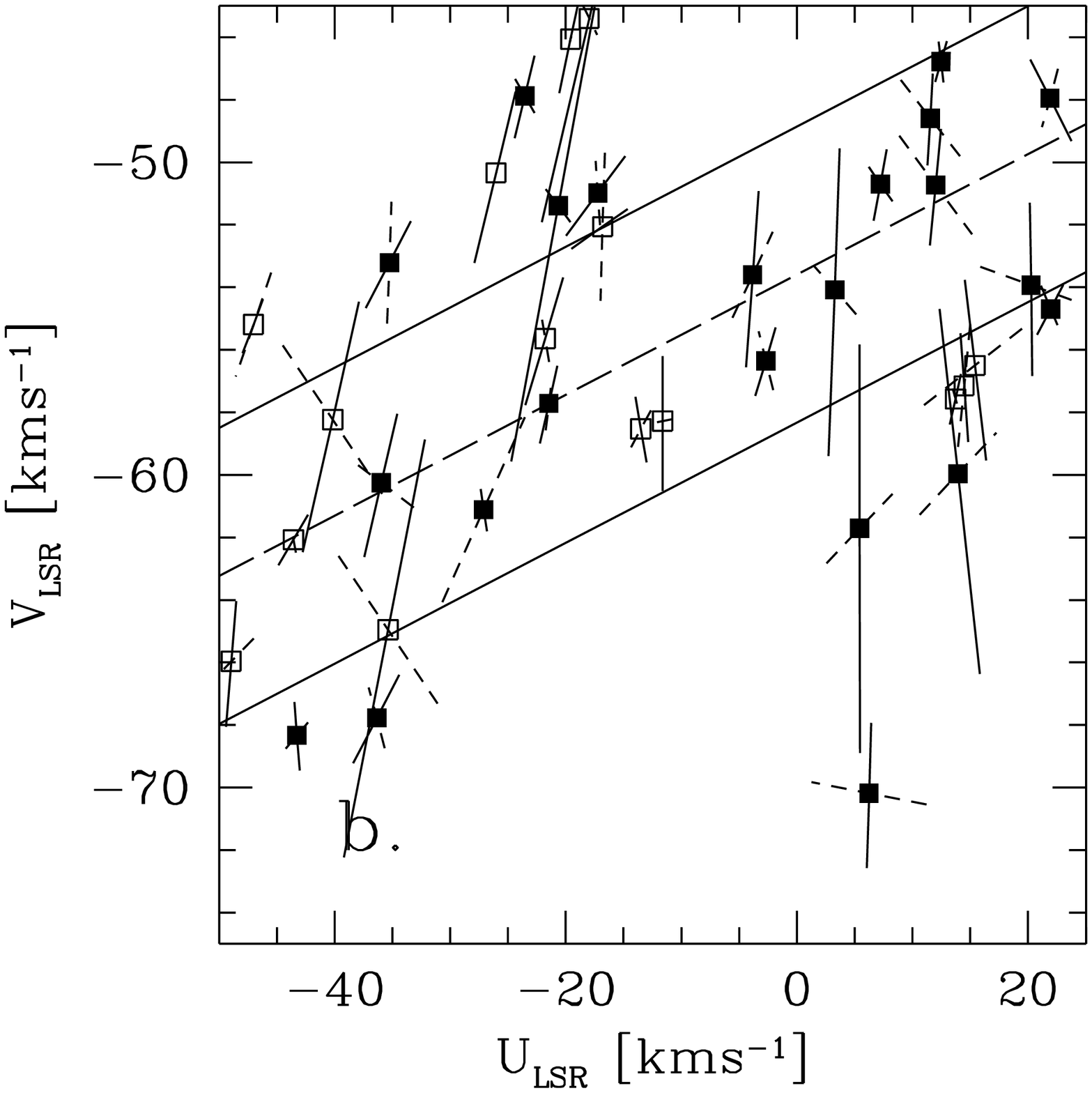}}
\hfill
\parbox[b]{55mm}{
\caption[]{Box 5 in the $UV$-plane with error-bars on the velocities arising
from errors in the parallaxes (full lines) and from errors in radial 
velocities (dashed lines). The tilted band outlines the boundaries of
our preferred HR1614 sample. {\bf a.}  Stars with $\sigma_{\pi} /\pi < 0.05$ 
and [Me/H] $>+0.1$ dex, {\bf b.}  Stars with $\sigma_{\pi} /\pi < 0.125$ and 
$0.0<$ [Me/H] $\leq +0.1$ dex, box, and [Me/H] $>+0.1$ dex, 
$\blacksquare$.}
\label{uvstr.fig}}
\end{figure*}

\subsection{Defining a new selection criterion}
\label{newsel.sect}

Is it possible to define a selection criterion in the $UV$-plane
for the moving group HR 1614 that is stricter than the boundaries of
box 5?  From Fig.  \ref{uvbox25.fig}  we find that 
metal-rich stars in box 5 predominantly fall along a diagonal band
going from the lower left to the upper right-hand corner. In order to
find out if this tilted structure is significant we select only stars
with small relative  errors in the parallaxes, $\sigma_{\pi} /\pi <
0.05$, and with [Me/H]$> 0.1$ dex. The latter selection is based on
the assumption that, if existing (as we assume in this section), the
HR 1614 group  should have a metallicity around $+0.2 $ dex. For
example the star  HR 1614 itself has [Fe/H] $=+0.28$ dex from high
resolution spectroscopy (Feltzing \& Gustafsson 1998). Fig.
\ref{uvstr.fig}a show the  $UV$-plane for these stars with error-bars
on the $U$ and $V$ velocities arising from errors in the parallaxes and
radial velocities.  Apart from one star (HIP 87116, which we
consider as belonging to the  general background and is excluded from
the  following discussion) the stars tend to fall along a diagonal
line. The star at the top ($U \simeq -24$ km~s$^{-1}$) could be
contamination from the U-anomaly (see text above),  but we leave it in
the sample. A simple least square fit to these data give $\Delta V /
\Delta U = +0.19$, Fig. \ref{uvstr.fig} a. We also  show lines
representing $\pm$ one rms. In Fig. \ref{uvstr.fig} b we  show all
stars in box 5 with  [Fe/H] $>0.1$ dex and  $\sigma_{\pi} /\pi <
0.05$ as well as all stars with $0.0<$ [Fe/H] $<0.1$ dex. As can be seen all of 
the stars with  [Fe/H] $>0.1$ dex
could be regarded as falling inside this tilted band if the errors on
the velocities are taken into account. This finding combined with our
simulations suggests that we have identified a metal-rich
co-moving stellar sample with a common origin in space and time. The
simulations indicate that the structures may not only be tilted but
also curved in the $UV$-plane, however, our data set is too small to
address that  question and we have to be satisfied with a ``straight
line''.

This selection procedure now helps us to define a 
new sample of probable HR 1614 moving group member stars
from our Str\"omgren catalogue, Table \ref{newsample.tab}.
Our new sample is limited by the fact that the calibration from
Schuster \& Nissen (1989) used to calculate [Me/H] is only valid for
dwarf stars with $0.32  \lesssim B-V \lesssim 1.0$. 
Because of these limitations K dwarf stars are excluded and
for example HR 1614 itself is not included although it perfectly obeys
our selection criteria in all other respects. However, see discussions
in Sect. \ref{sect.met} and \ref{eggen.sect}.

We thus propose that future searches for member stars of the HR 1614
moving groups should be directed to the area in the $UV$-plane defined by
our tilted band. 

\subsection{A new estimate of the age of the moving group HR 1614}

\begin{figure}
\resizebox{\hsize}{!}{\includegraphics{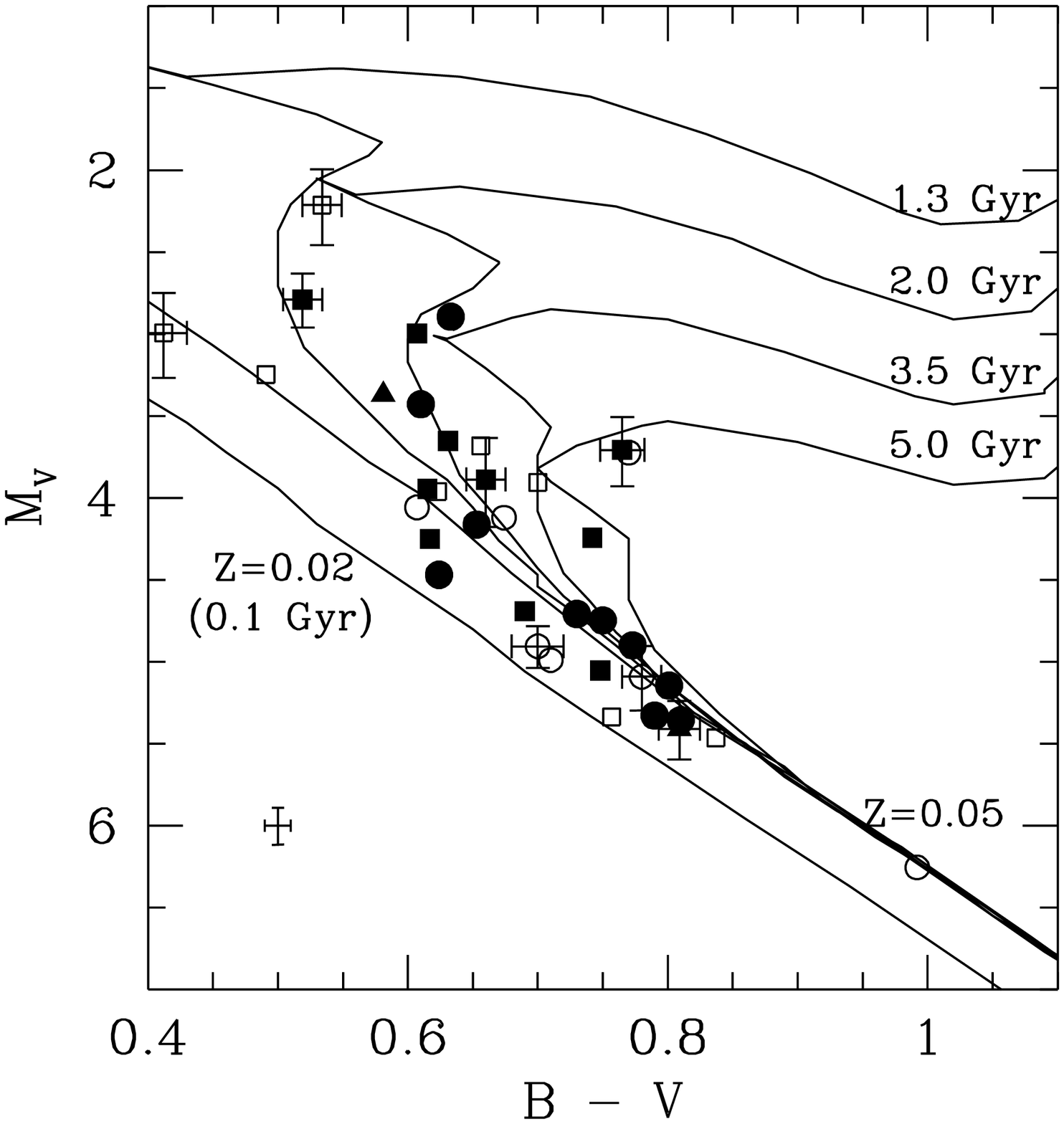}}
\caption[]{The HR-diagram for box 5 for stars with [Me/H]$>0.0$ dex.
All stars have  $\sigma_{\pi} /\pi < 0.125$ and all filled symbols
refers to stars with [Me/H] $>0.1$ while open symbols to stars with
 $0.0<$ [Me/H] $<0.1$. Typical $1 \sigma$ 
error-bars on $M_V$ and $B-V$ are 
shown in the lower left corner. For stars with larger error-bars
we explicitly draw those. 
$\bullet$ refers to stars inside the tilted strip in Fig. 
\ref{uvstr.fig} a with [Me/H] $>0.1$ dex and  $\sigma_{\pi} /\pi 
< 0.05$. $\blacktriangle$
the remaining metal-rich stars inside the strip but with larger errors.
$\blacksquare$ stars with [Me/H]$>0.1$ dex  and outside the strip. 
$\bigcirc$ stars with [Me/H] $<0.1$ and inside the strip, and finally
box stars outside the strip and with [Me/H] $<0.1$. The ZAMS are
from   Bertelli et al. (1994)
for $Z=0.02$ and $0.05$ 
which correspond to [Fe/H] = 0.0 and 0.4 dex. For $Z=0.05$ we also show
the isochrones with ages as indicated.}
\label{hrbox5age.fig}
\end{figure}

We are now in a position to determine a new age estimate of the 
HR 1614 moving group from the HR-diagram of stars with Str\"omgren 
metallicities and based on the selection described in the 
previous section.
Thus from
Fig. \ref{hrbox5age.fig} we derive an age estimate  of 2 Gyr using
the Bertelli et al. (1994) isochrones.  

The iron content [Fe/H] of the Z = 0.05 isochrone from Bertelli et
al. (1994)  which is found to be [Fe/H] =  +0.40 dex from their relation (11),
is higher then the observed one for our sample of HR 1614 member stars
which is $+0.19\pm0.06$.  This discrepancy could be due to an error in
the colour transformation from the $\log T_{\rm e},M_{\rm bol}$ to the
$B-V,M_{\rm V}$ plane where a metal-rich model  atmosphere is used
compared to the solar metallicity case which is much  better
constrained. Other possible error sources are the  assumed $\Delta Y /
\Delta Z$, where a lower value would give a lower  metallicity for the
isochrone and the assumed scaled solar abundances. If other elements
than iron is over-abundant this would lower the iron content for the
same assumed metallicity. 

In an attempt to estimate an
error on this age estimate we have compared the  the Bertelli et
al. (1994) isochrones with those from Chaboyer et al.
(1999). Their metal-rich isochrones present excellent fits  to
observations from solar metallicity (M67) to the very high metallicity 
of [Fe/H] $=+0.4$ for NGC 6791,
both of which are older than the HR 1614 moving group.  For solar
metallicity the Bertelli et al. (1994)  and Chaboyer et al.
(1999)  isochrones are very similar up to the sub-giant region.  
The two sets of  isochrones
agree well for ages applicable to NGC 6791. Chaboyer at al. (1999)
find this open cluster to be 8 Gyr. Furthermore, and most importantly
here, they also investigate the effect of changing  $\Delta Y / \Delta
Z$. Using values between 1 and 3 they find the impact on the  derived
age to be small.  Bertelli et al. (1994) use  $\Delta Y / \Delta
Z=2.5$.  It is quite likely that the HR 1614 moving group has a
$\Delta Y / \Delta Z \neq 2.5$. From our
comparison of the isochrones we conclude that errors on the  derived
age, from the possibility that our stars have  $\Delta Y / \Delta Z
\neq 2.5$, is less than 1 Gyr.

The conclusion by Hufnagel \& Smith (1994) that the age of the HR 1614
group  is $\geq$3 Gyr based on the MgII chromospheric emission index
is consistent with our isochrone determined age of 2 Gyr in the light
of the work  by Rocha-Pinto \& Maciel (1998). They show that the
strength of the  chromospheric emission has a strong dependence of the
metallicity of the  stars, with metal-rich stars having higher
chromospheric ages than isochrone  ages. Using their relation between
the error in the chromospheric age and the  iron abundance, for a
assumed [Fe/H] of +0.20 for HR 1614, a chromospheric age  of 3 Gyr
corresponds to a isochrone age of 1 Gyr.

\section{The metallicity of the HR 1614 moving group}

\begin{figure*}
\resizebox{12cm}{!}{\includegraphics{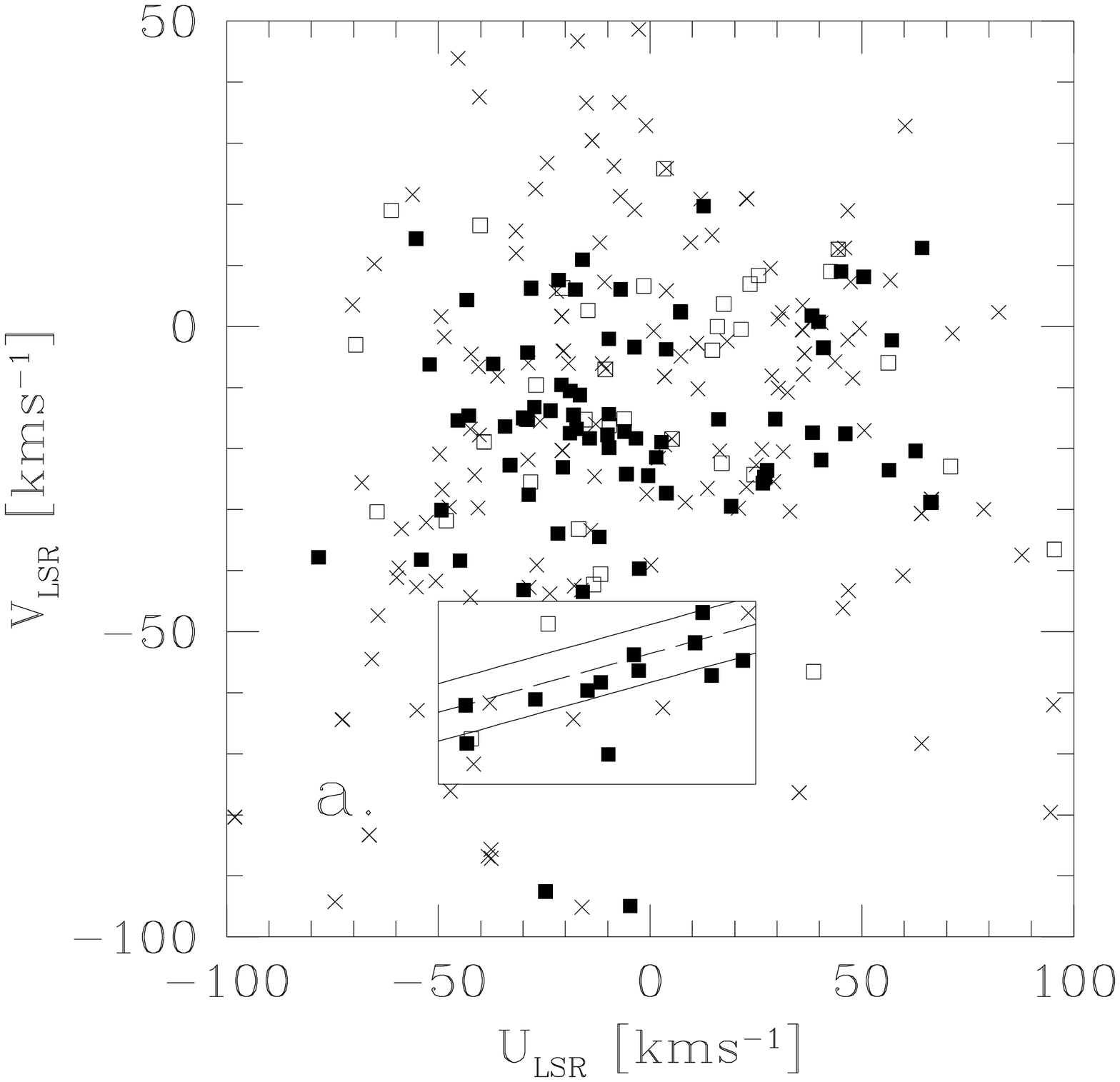}
\includegraphics{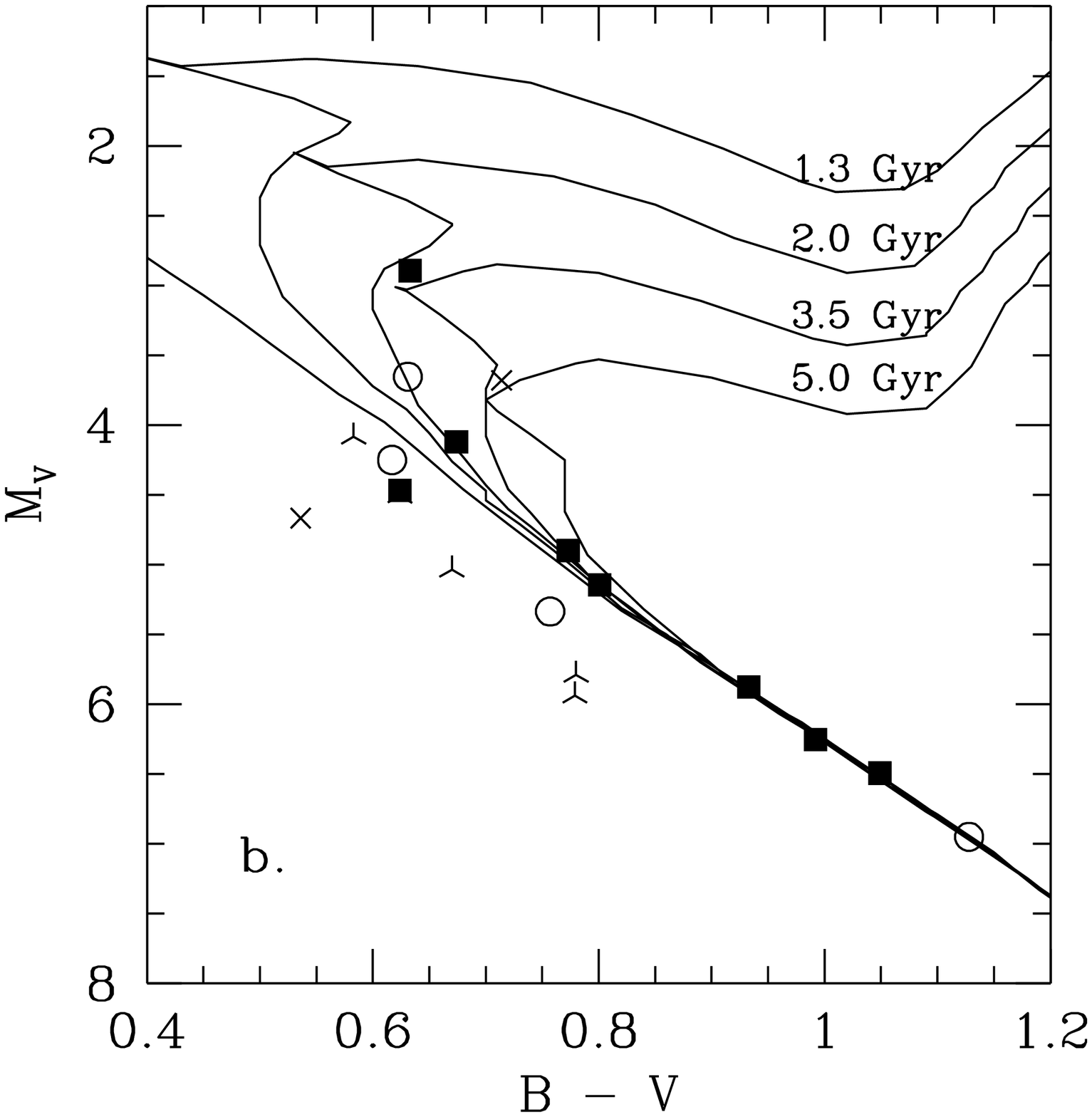}}
\hfill
\parbox[b]{55mm}{
\caption[]{{\bf a.} $UV$-plane plot of the spectroscopic catalogue. $\times$
represents stars with [Fe/H] $< -0.05$ dex, box stars with
$-0.05 \leq$[Fe/H]  $< 0.05$, and $\blacksquare$ stars with  $0.05
\geq$[Fe/H]. Box 5 is marked and inside the box
we reproduce the lines from Fig. \ref{uvstr.fig}. 
The HR-diagram in {\bf b.} shows the stars inside the 
tilted strip, $\blacksquare$ with [Fe/H] $\geq 0.05$ dex,
 $\times$ $ < 0.05$ dex,
and stars outside the strip,  $\bigcirc$ with [Fe/H] $\geq 0.05$ dex, 
triangle crosses [Fe/H] $< 0.05$ dex }
\label{uv.starcat.fig}}
\end{figure*}

We now proceed to determine the metallicity of the HR 1614 moving group
from [Fe/H] measurements found in the literature.

\subsection{The spectroscopic catalogue}
\label{catalogues.sect}

The basis for our spectroscopic catalogue is the kinematic catalogue described
in Sect. \ref{search.sect}
supplemented by iron abundances as
well as other elemental abundances  from the catalogues by   Fuhrmann
(1998), Favata et al. (1997), Edvardsson et al. (1993), and Feltzing
\& Gustafsson (1998). 

These studies were  chosen because they
all have large samples of stars, high-resolution has been used, they
all have small internal errors, and a comparison of the several stars
in common between the studies reveals a  high internal consistency
between the studies, $\sigma$([Fe/H])$= \pm 0.06 $ dex. All
kinematical data have been recalculated using the new parallaxes and
proper motions from Hipparcos.

The Edvardsson et al. (1993) sample was selected in order to study F
and early G main sequence stars, which were evenly distributed in
metallicity between $-1.0$ dex and $+0.3$ dex, and slightly evolved
off the ZAMS so that  ages could be estimated. The selection of stars
was done from Olsen (1988) which includes almost all F and early G
stars brighter than $V \simeq 8.3$.  The stars in Feltzing \&
Gustafsson (1998) were selected from the same photometric
catalogue with updates in order to study the metal-rich population of
disk dwarf stars on solar orbits but also of stars with fairly
eccentric orbits and a subsample was selected to have
$V_{\rm LSR}  < - 50$ km~s$^{-1}$ and/or total space 
velocity $> 60$ km~s$^{-1}$. 
We may therefore expect  to find most probable HR
1614 moving group members in this study. Note that in the selection 
of stars for these two studies no attempt was made to cover the probable 
member stars of the HR 1614 moving group.

Favata et al. (1997) provide a volume limited sample of G and K dwarf
stars drawn from the Gliese catalogue. Fuhrmann (1998) provide iron
abundances for 50 nearby F and G type stars on the main-sequence,
turn-off, and sub-giant branch. 

Comparing [Fe/H] from Edvardsson et al. (1993) and Feltzing \&
Gustafsson (1998)  with [Me/H] derived 
using the calibration by Schuster \& Nissen (1989) we find that
${\rm [Fe/H]} - {\rm [Me/H]}= +0.015 \pm0.076$. 

\subsection{The metallicity}
\label{sect.met}

We now apply the $UV$-selection criterion to our spectroscopic
catalogue.  The resulting HR diagram and also the $UV$-plot for the full
catalogue are shown in Fig. \ref{uv.starcat.fig}.

As noted in Sect. \ref{newsel.sect} our final sample of probable
member stars presented in Table \ref{newsample.tab} only contained
stars which fell inside the calibration by Schuster \& Nissen (1989),
however, for example HR 1614 itself would fall outside this
calibration but does indeed fulfill the  selection criteria. We thus
include two K dwarf stars in our final sample of probable member stars
on the basis of their spectroscopically measured  metallicities.

For the 8 stars with [Fe/H]$>0.05$ dex  $<$[Fe/H]$>$ $ = 0.19 \pm
0.06$ using the [Fe/H] for HR 1614 from Feltzing \& Gustafsson
(1998). For the 11 rejected stars (i.e. both stars inside the strip with
[Fe/H] $<0.05$ dex and stars outside the strip)  we find $<$[Fe/H]$>$
$ = -0.19 \pm 0.38$. The three that fall just outside our selection
criterion in the $UV$-plane (HIP 22336, HIP 26437, HIP 70470) have [Fe/H]
= 0.14, 0.15, 0.06 dex. Including them into the sample we arrive at
$0.17 \pm 0.06$ dex.

Obviously the spectroscopic catalogue is subject to selection biases.
However, none of the investigations included in the catalogue have
been biased {\sl against} metal-rich stars in this part of the
$UV$-plane.  In fact Feltzing \& Gustafsson (1998) did actively include
metal-rich stars with large negative $V_{\rm LSR}$. Note though that no
attempt was made in their investigation to preferentially include HR 1614 
members. We
may  thus expect to sample the metal-rich stars well, while the
metal-poor stars are less well sampled.

In conclusion we find that the HR 1614 moving group has a metallicity,
determined from high resolution spectroscopy, of $0.17-0.19 \pm 0.06$
dex depending on the exact selection in the $UV$-plane. 

\begin{table*}
\caption[]{Our new sample of HR1614 stars. Columns 1 and 2  give the
Hipparcos and HD numbers for the stars, column three the parallaxes, column 
four the apparent $V$ magnitudes, column five the $B-V$ 
colour, column six to eight give the
new U, V, and W velocities, column nine the [Me/H] metallicities calculated 
from Str\"omgren photometry
using  the calibration by Schuster \& Nissen (1989) and column nine the
spectroscopically determined iron abundances. All stars 
have $\sigma_{\rm B-V} < 0.02$, for most stars it is less than 0.01,
and $\sigma_{\pi} / \pi < 0.125$. In the column 
headed Sample $\bullet$ refers to stars inside the tilted strip in Fig. 
\ref{uvstr.fig} a with [Me/H] $>0.1$ dex and  $\sigma_{\pi} /\pi 
< 0.05$. $\blacktriangle$
the remaining metal-rich stars inside the strip but with larger errors.
$\blacksquare$ stars with [Me/H] $>0.1$ dex  and outside the strip. 
$\bigcirc$ stars with [Me/H] $\leq0.1$ and inside the strip, and finally
box stars outside the strip and with [Me/H] $\leq0.1$. In the last column
in the first two parts of the table we indicate which stars are in common with
Eggen (1992, 1998b) sample. }
\begin{tabular}{rrrrrrrrrrcr}
\hline\noalign{\smallskip}
HIP & HD & \multicolumn{1}{c}{$\pi$} (mas) & \multicolumn{1}{c}{V} & 
\multicolumn{1}{c}{$B-V$} & \multicolumn{1}{c}{$U_{\rm LSR}$}& 
\multicolumn{1}{c}{$V_{\rm LSR}$} & \multicolumn{1}{c}{$W_{\rm LSR}$} 
& [Me/H] &[Fe/H] & Sample \\
\noalign{\smallskip}
\hline\noalign{\smallskip}
  9353 &  12235 & 32.18$\pm$0.96   & 5.89 &0.610 &  11.55 & -48.59 &  12.43 & 0.21 & &$\bullet$ & e\\
 10505&	 13825 &  37.87$\pm$0.96 & 6.80&  0.690&  -23.54&  -47.88&   8.17 & 0.15  & & $\blacksquare$\\
 10599 &  13997 & 29.35$\pm$1.13  & 7.99 &0.790 &  12.00 & -50.73 &  13.64 & 0.14 & &$\bullet$& e\\
 15934 & 275241 & 13.56$\pm$1.22  & 9.43 &0.780 & -35.37 & -64.95 &  -2.66 & 0.02 & & $\circ$\\
 16467 &  21727 & 18.58$\pm$1.09  & 8.56 &0.700 & -40.17 & -58.23 &  24.64 & 0.01 & & $\circ$\\
 17960 &  24040 & 21.50$\pm$1.03  & 7.50 &0.653 &  20.26 & -53.93 &  -7.57 & 0.21 & & $\bullet$\\
 22336 &  30562 & 37.73$\pm$0.89  & 5.77 &0.631 & -43.26 & -68.33 & -14.98 & 0.11 &0.19 &$\blacksquare$ \\
 22940 &  31452 & 25.50$\pm$1.27  & 8.43 &0.837 &  15.40 & -56.50 &  -3.05 & 0.03 & & $\circ$& \\
 25094 &  34575 & 34.00$\pm$0.86  & 7.09 &0.750 &   7.23 & -50.70 &   7.00 & 0.18 & & $\bullet$\\
 26437 &  36130 & 19.96$\pm$0.73  & 7.75 &0.617 &  21.95 & -54.69 & -37.55 & 0.26 &0.15 & $\blacksquare$\\
 26834 &  37986 & 36.05$\pm$0.92  & 7.36 &0.801 & -27.11 & -61.11 &   3.02 & 0.17 &0.27 & $\bullet$\\
 28179 &  41158 &  9.16$\pm$0.69  & 7.98 &0.519 & -17.23 & -50.98 &   2.57 & 0.15 & & $\blacksquare$\\
 46325 &  81505 & 11.52$\pm$1.12  & 8.40 &0.765 &  13.93 & -59.96 &  16.14 & 0.24 & & $\blacksquare$\\
 49060 &  86680 & 10.03$\pm$1.00  & 7.99 &0.607 &   5.43 & -61.71 &  -4.23 & 0.26 & & $\blacksquare$\\
 51257 &  90711 & 31.12$\pm$0.94  & 7.89 &0.810 & -21.46 & -57.71 & -25.27 & 0.17 & &$\bullet$&e\\
 52990&	 93932 &  19.18$\pm$0.88 & 7.53&  0.615&  -35.26&  -53.21& -13.40 & 0.16 & & $\blacksquare$\\
 53537 &  94835 & 20.22$\pm$1.01  & 7.94 &0.624 &  -3.83 & -53.60 &  -7.10 & 0.15 &0.13 & $\bullet$\\
 54632 &  96511 & 22.43$\pm$0.56  & 7.15 &0.700 &  13.74 & -57.57 &   1.04 & 0.09 & & $\circ$& \\
 60729 & 108309 & 37.50$\pm$0.72  & 6.25 &0.674 & -43.56 & -62.07 &  -3.35 & 0.07 &0.10 & $\circ$\\
 60829 & 108523 & 21.67$\pm$0.93  & 8.31 &0.710 & -21.78 & -55.64 &  -3.14 & 0.02 & & $\circ$\\
 61435&	109542 &   9.01$\pm$1.07 & 8.22&  0.412&  -17.97&  -45.40& -10.78 & 0.05 & & box\\
 62857&	112001 &  18.22$\pm$0.90 & 7.66&  0.623&  -26.03&  -50.34&  -4.78 & 0.05 & & box\\
 63033 & 112164 & 25.17$\pm$0.76  & 5.89 &0.633 &  -2.68 & -56.35 & -23.38 & 0.23 &0.24 & $\bullet$ & e\\
 67195 & 120005 &  22.26$\pm$0.75  & 6.51 &  0.491 & -19.58 & -46.06 &  11.65 &    0.07 & & box\\
 65036 & 115585 & 23.05$\pm$0.70  & 7.43 &0.742 & -36.35 & -67.77 &   4.93 & 0.17 & & $\blacksquare$\\
 70470 & 126511 & 24.84$\pm$0.95  & 8.36 &0.757 &  14.47 & -57.15 & -13.97 & 0.07 &0.06 & $\circ$\\
 75266 & 136834 & 39.35$\pm$1.37  & 8.28 &0.992 & -11.61 & -58.28 & -10.05 & 0.05 &0.16 & $\circ$& e\\
 79240 & 144899 &  9.63$\pm$1.20  & 8.97 &0.660 & -20.65 & -51.39 &   7.32 & 0.12 & &$\blacksquare$ \\
 83435 & 154160 & 27.58$\pm$0.79  & 6.52 &0.770 & -13.50 & -58.53 &  -8.86 & 0.04 & &$\circ$ \\
 87116 & 161612 & 37.19$\pm$1.15  & 7.20 &0.748 &   6.24 &  -70.18&  -7.76 & 0.16 & & $\blacksquare$\\
 91283 & 172085 & 14.86$\pm$0.80  & 7.51 &0.581 &  21.90 & -47.95 &  -2.18 & 0.10 & & $\blacktriangle$&e\\
 94797 & 181358 &  6.32$\pm$0.67  & 8.21 &0.534 & -16.81 & -52.06 &  13.79 & 0.06 & & box\\
102018 & 196800 & 23.42$\pm$0.89  & 7.21 &0.607 & -48.95 & -65.96 &  -2.84 & 0.05 & & $\circ$ \\
102393 & 197623 &  16.84$\pm$1.11 & 7.55&  0.656&  -47.02&  -55.19&   -3.02& 0.04 & & box\\
109378 & 210277 & 46.97$\pm$0.79  & 6.54 &0.773 &  12.46 & -46.78 &   3.09 & 0.16 &0.22 &$\bullet$&e\\
110843 & 212708 & 27.90$\pm$0.96  & 7.48 &0.730 & -35.96 & -60.25 &  -2.93 & 0.13 & & $\bullet$\\
116554 & 222013 & 17.29$\pm$1.42  & 9.22 &0.809 &   3.27 & -54.08 &  17.25 & 0.12 & &$\blacktriangle$&e\\
\noalign{\smallskip}
\hline\noalign{\smallskip}
\noalign{\smallskip}
\multicolumn{10}{c}{K dwarf stars with spectroscopically determined [Fe/H]}\\
\noalign{\smallskip}
\hline\noalign{\smallskip}
\noalign{\smallskip}
 13513 & 18168 & 33.83$\pm$0.98  &  8.23 & 0.933 & -14.75&  -59.61 &  39.00  &  & 0.18 &&   \\
 23311 & 32147 & 113.46$\pm$0.82 &  6.22 & 1.049 &  10.65&  -51.83 &  -6.33  &   &0.22 & &e\\ 
\noalign{\smallskip}
\hline\noalign{\smallskip}
\noalign{\smallskip}
\multicolumn{10}{c}{Additional stars from the Eggen (1992, 1998b) sample} &\\
\noalign{\smallskip}
\hline\noalign{\smallskip}
\noalign{\smallskip}
6762  &   8828 &  33.59$\pm$ 0.99 &  7.96 &  0.738 &   -4.09 &  -53.95 &  -22.13 & & &\\  
21932  & 285968 & 106.16$\pm$ 2.51 &  9.95 &  1.523 &  -15.49 &  -52.37 &   -8.60\\  
41136  &        &  18.99$\pm$ 1.92 & 10.04 &  0.960 &    4.42 &  -49.58 &   -9.00\\   
94225  & 178445 &  49.84$\pm$ 1.51 &  9.36 &  1.318 &   -2.73 &  -53.10 &   -4.50\\   
110996  & 213042 &  64.74$\pm$ 1.07 &  7.65 &  1.080 &   16.55 &  -54.51 &   -9.70\\ 
116970  & 222655 &  11.84$\pm$ 1.43 &  9.57 &  0.760 &    9.87 &  -55.64 &  -15.42\\   
\hline
\end{tabular}
\label{newsample.tab}
\end{table*}

\section{The Eggen sample revisited}
\label{eggen.sect}

\begin{figure*}
\resizebox{\hsize}{!}{\includegraphics{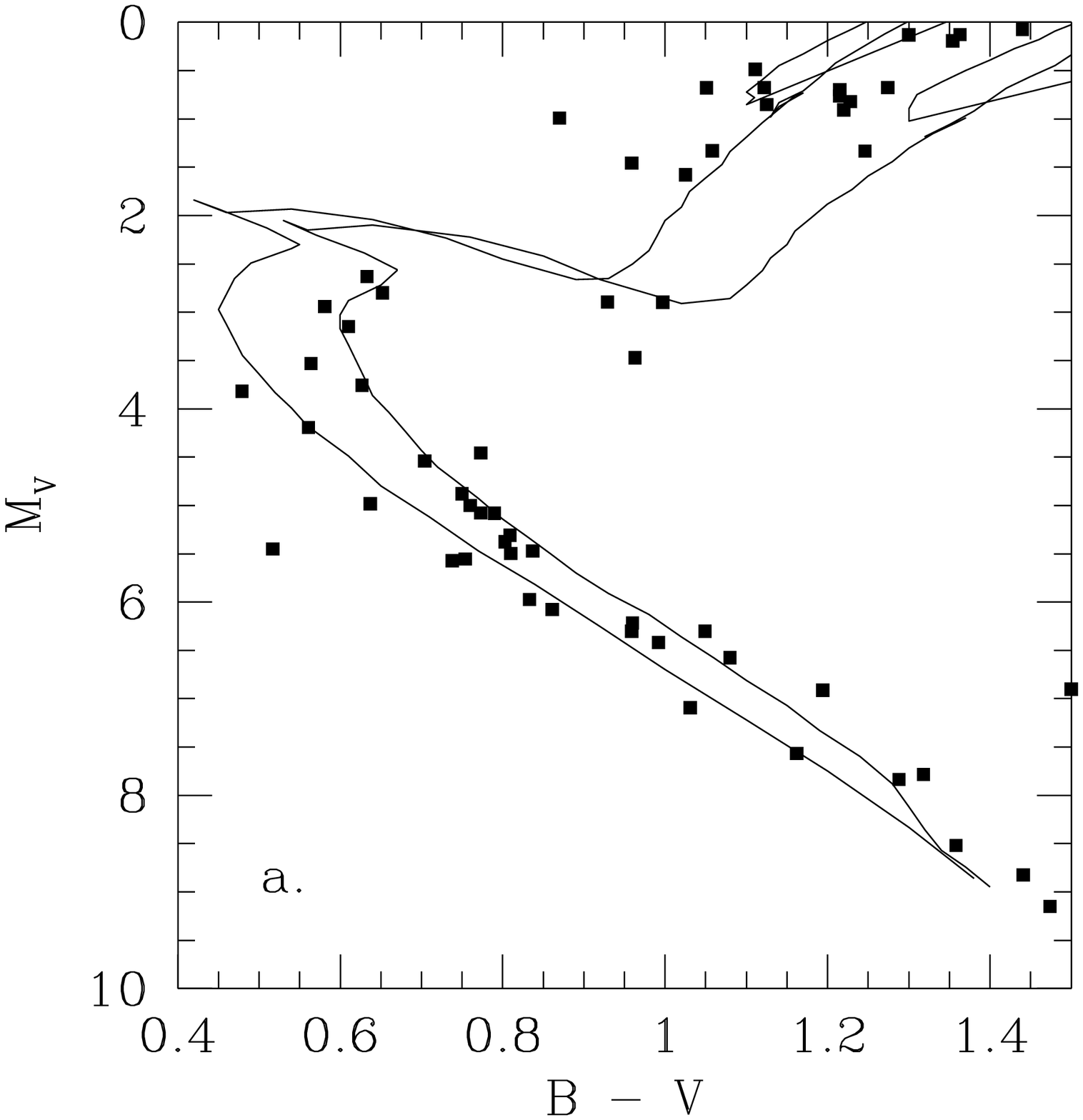}
\includegraphics{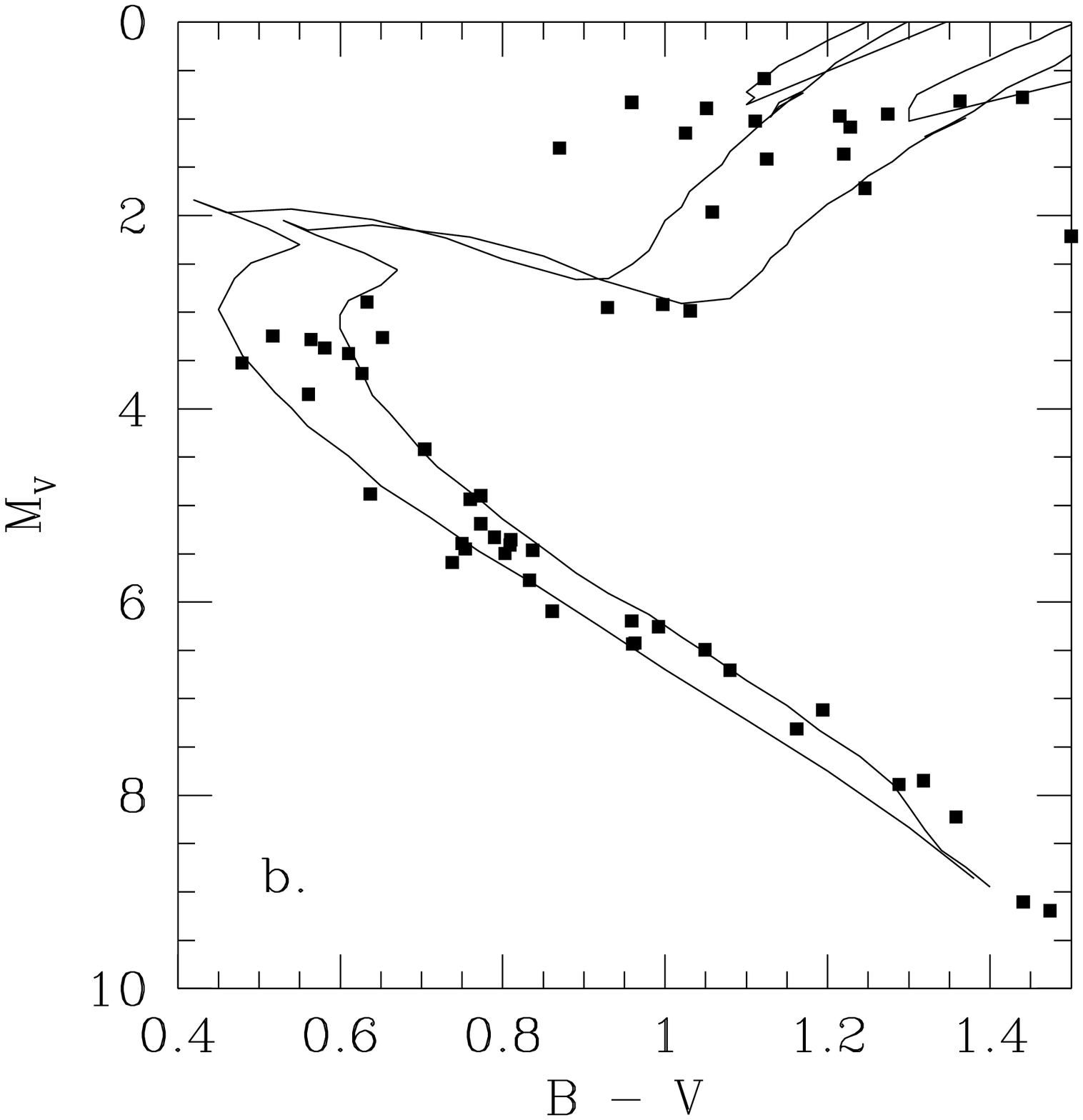}\includegraphics{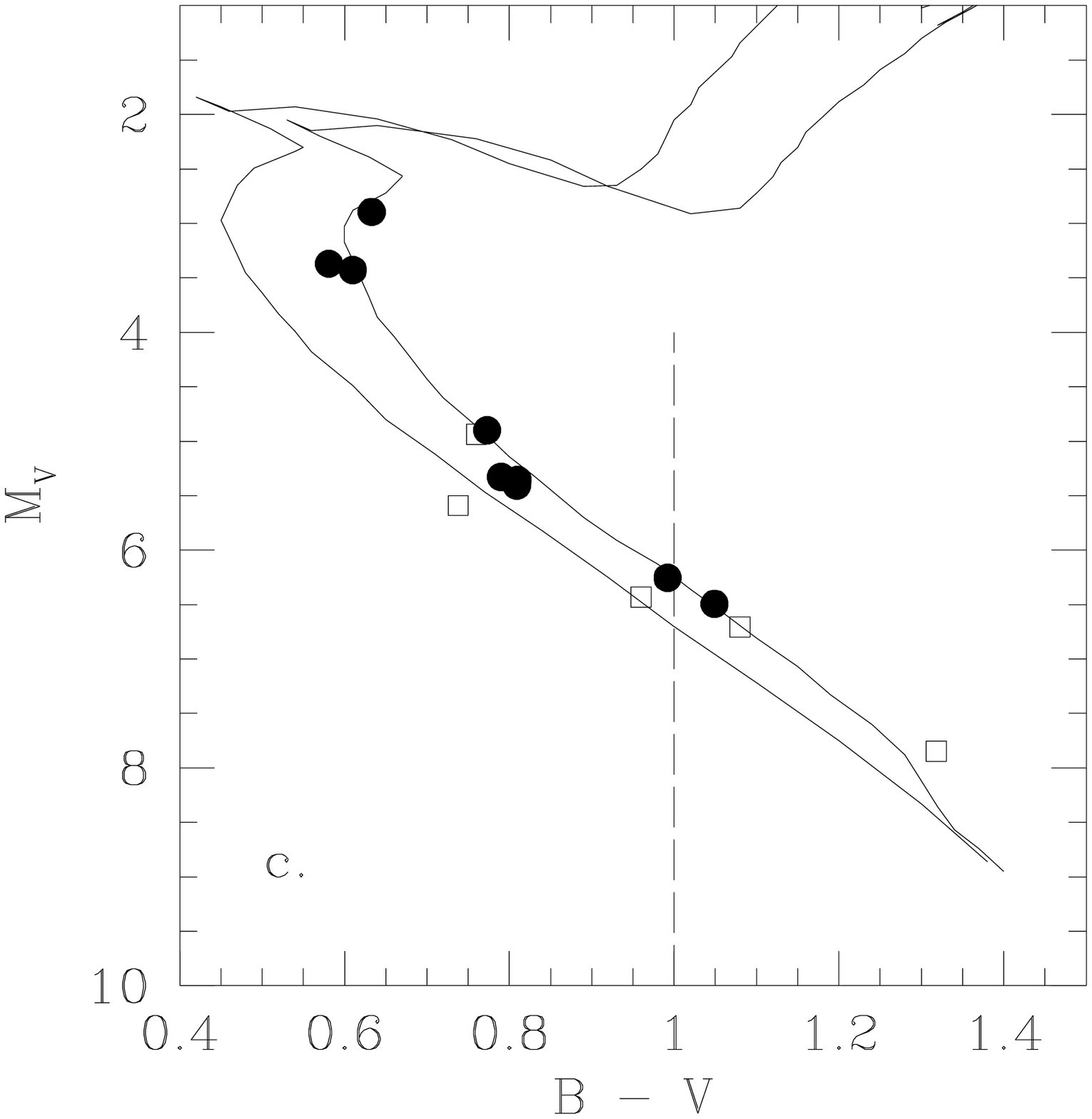}}
\caption[]{Eggen (1998b) HR-diagram. {\bf a.} Using his cluster parallaxes,
{\bf b.} using Hipparcos parallaxes, and {c.} using Hipparcos parallaxes
and imposing our selection criteria in the $UV$-plane. $\bullet$ 
denotes stars that were found independently in our search,
box stars obeying the selection criteria and from which
known binaries and stars with too low metallicities (when the metallicity
is known) has been removed. The limit of the calibration by Schuster \&
Nissen (1989) is marked with a dashed line. }
\label{hreggen.fig}
\end{figure*}

Having defined a new selection criterion for membership in the
$UV$-plane we now revisit the Eggen (1992, 1998b) sample of HR 1614 moving
group stars.  Fig. \ref{hreggen.fig} show the sample of stars that
Eggen (1998b)  indicate as probable members of the HR 1614 moving group.  
In  Fig. \ref{hreggen.fig}a we use his cluster 
parallaxes, in  Fig. \ref{hreggen.fig}b the Hipparcos
parallaxes.  The use of the cluster parallaxes forces the stars onto more
or less one isochrone. Eggen (1998b) does not discuss the apparent
metallicity spread in Fig. \ref{hreggen.fig}a. 
We note especially that the turn-off region,
crucial for age determination, shows a larger scatter in
Fig. \ref{hreggen.fig}b as compared to Fig. \ref{hreggen.fig}a.

When we  apply our $UV$-selection criteria to Eggen's sample (with
re-calculated velocities and using the Hipparcos parallaxes) a total
of 21 stars with $M_V>2$ falls inside the band in Fig. \ref{uvstr.fig}. 8 of
these  are in common with our selection from the Str\"omgren
catalogue. These stars are marked in the first part of Table
\ref{newsample.tab}. Of the remaining 13 stars one is HR 1614 and is
already included in our sample on the basis of its [Fe/H] measured
from high resolution spectroscopy.  The two stars HIP 11575 and HIP
35872 appears in our Str\"omgren catalogue and have [Me/H]$= -0.12$
and $-0.09$, respectively, disqualifying them as probable member stars
of the HR 1614 moving group.  Four more stars were excluded due to
binarity, HIP 7143 is a spectroscopic binary, HIP 94570 and HIP 96037 were
resolved as binary systems by Hipparcos and HIP 112426 have an
accelerated solution, probably due to orbital motion (ESA 1997).

This leaves, after having  considered binarity and metallicity when
information is available, an additional 6 stars from Eggen's (1992,
1998b) sample to be included in our final sample of probable member
stars of the HR 1614 moving group. The stars are detailed in the final
part of Table \ref{newsample.tab}.

The HR-diagram of the 15 stars from Eggen (1992, 1998b) thus
full-filling our criteria are plotted  in Fig. \ref{hreggen.fig}c.

\section{Discussion}

As already touched upon, the HR-diagram of the HR 1614 group is 
``contaminated'' 
by stars from the general field and possibly also by stars from other 
moving groups and open clusters. 

The general background of stars in the $UV$-plane using the Hipparcos
results have been studied by  Dehnen (1999) and Raboud et al. (1998).
Dehnen (1999) shows that there exists an over-density of stars in the
$UV$-plane centered on $U=-20$ and $V=-45$ km~s$^{-1}$. He associates this 
stellar
over-density with stars thrown out from the inner disk  by the galactic
bar. In particular stars close to the outer Lindblad resonance
are susceptible to this and the phenomenon a clear indication of the
non-axisymmetry  of the galactic potential. Thus we should expect to
see a generally more metal-rich  stellar population present in the
upper left half of our boxes in Fig.  \ref{hrboxes.fig}. We do indeed
do so, however, we also note that the  contamination is mainly in box
1 and 2 and does not effect box 5. This is further born out by
the $UV$-plots using the Str\"omgren sample, Sect. \ref{search.sect}.

The moving group Wolf 630 is present in the upper right hand corner of
box 5 ($-50 < U< +25$ and $-75 < V < -45$ km~s$^{-1}$, see also Fig. 1 in
Skuljan et al. 1997 for the position of Wolf 630). 
However, since the Wolf 630 moving group has
a mean [Fe/H] of $\sim -0.14$ dex (Boesgaard \& Friel 1990) almost
none of the  stars would remain in our Str\"omgren sample.
 Schuster \& Nissen (1989)
estimated the scatter of their calibration 
to 0.16 dex. When we compare the estimated
metallicities with spectroscopic abundances the differences are much
smaller, around 0.06 dex. Thus we may expect these stars to be removed
also in the Str\"omgren sample. In the spectroscopic sample all
 would be removed.  We conclude that the contamination of our
sample by stars from the  Wolf 630 moving group is negligible.

Eggen (1998a) suggested that the open cluster NGC 6791 might be part
of the HR 1614 moving group. This is ruled out by two recent studies
by Tripico et al. (1995) and Chaboyer et al. (1999) which both
agree that the open cluster has an age around $8-10$ Gyr.  This means that
even if we allow for the uncertainties in our age  determination due
to uncertainties in the modeling of stellar evolutionary tracks the
age of the HR 1614 moving group is $>$ 5 Gyr younger than  the old open
cluster NGC 6791. Also, the metallicity of NGC 6791 is almost 0.2 dex 
higher than that of the HR 1614 moving group. Combining
these facts the association of the HR 1614 moving group with NGC 6791 
must be incorrect.

Our sample provides a sample for future studies of the abundance
profiles in old moving groups to address the question whether or 
not todays field stars originate in stellar clusters subsequently
dissolved. 
It is at  the higher metallicities, i.e. ${\rm [Fe/H]} \geq 0.1 $ dex,
that differences in star formation histories will manifest themselves
in the abundance ratios (e.g. Pagel 1997,  fig 8.6, and Matteucci 1991). If
metal-rich stars  in the field originate in clusters then  abundance
ratios for  stars in a metal-rich moving group should be identical to
those of the field stars in the solar neighbourhood. On the other
hand if the moving groups are not the source of the field stars then
it is most likely that their star formation rate was different and
thus the resulting elemental abundances will be different. If the star
formation rate was more rapid in the cluster than what is typical for
the places where the metal-rich stars, now in the solar  neighborhood,
were born then the [$\alpha$/Fe] will be  larger at say
[Fe/H]$\sim0.2$ dex for the group stars than for the field stars. If
on the other hand the star formation rate was lower [$\alpha$/Fe] will
be smaller, Pagel (1997 Fig. 8.6).
 
\section{Summary}

Utilizing the new possibilities given by the Hipparcos mission we
perform, for the first time, an unbiased search in the $UV$-plane of all
stars with measured radial velocities and find an over-density of
stars more metal-rich than the Sun close to the $U$ and $V$ values
associated with the moving group HR 1614 (Eggen 1998b).  
Supported by dynamical simulations we find that old moving groups in
fact does exist, at least the HR 1614 moving group.  The selection
criterion for the HR 1614 moving group in the $UV$-space is further
refined using metallicities derived from Str\"omgren photometry.
This new criterion is applied to the large catalogue of all Hipparcos
stars with measured radial velocities and Str\"omgren photometry
resulting in a new determination of the probable age of the moving
group. 

We derive an age of 2 Gyr using the Bertelli et al. (1994) isochrones
and a iron abundance of $+0.19 \pm 0.06$ dex using available data,
from high resolution spectroscopy, in the literature.

We also provide a comparison of [Me/H] derived from Str\"omgren
photometry of the Hauk \& Mermilliond catalogue (1998) using the Schuster \&
Nissen (1989) calibration. The comparison shows that the Str\"omgren
[Me/H] and [Fe/H] for F and G type dwarf and sub-giant stars are in
extremely good agreement with ${\rm [Fe/H]} - {\rm [Me/H]}= +0.015
\pm0.076$.

The sample of bright main-sequence and turn-off stars that are
probable members of the moving group  HR 1614 is  presented in Table
\ref{newsample.tab}. They make up a sample for further investigations
into the abundance profile of this old, metal-rich moving group.

\begin{acknowledgements}
This work has made use of the SIMBAD on-line database search facility
maintained by the CDS, Strasbourg. JH acknowledges the financial support 
of the Swedish National Space Board. The Royal Physiographic Society in Lund is 
thanked for providing funding for computer facilities. The anonymous 
referee is thanked for providing contructive remarks that lead to a 
revision and considerable improvement of the paper. 
\end{acknowledgements}

\end{document}